\documentclass [11pt,a4paper] {article}

\usepackage {amssymb, amsthm, mathrsfs}
\usepackage {hyperref}
\usepackage {preview}
\usepackage{fullpage}

\newtheorem {theorem}{Proposition}
\newtheorem {lemma}{Lemma}

\def\bsk {\vspace {1.5 cm}}
\def\msk {\vspace {1.0 cm}}
\def\ssk {\vspace {0.2cm}}
\def\ni {\noindent}

\begin {document}

\bsk
\centerline {\huge {Solvable Models Of Infrared Gupta-Bleuler}}
\ssk
\centerline {\huge {Quantum Electrodynamics}}

\ssk

\msk

\centerline {\Large {Simone Zerella}*}
\centerline {\emph {Dipartimento di Fisica dell'Universit\`a, I-56126 Pisa, Italy}}

\ssk

\bsk\ni

\abstract{
\ni
Hamiltonian models based on two different infrared approximations are studied in order to obtain 
an explicit comparison with the standard analysis of the infrared contributions, occurring in the 
relativistically covariant perturbative formulation of Quantum Electrodynamics.\\ 
M\"{o}ller operators, preserving respectively the Hilbert scalar product, for the Coulomb-gauge models, 
and an indefinite metric, for the models formulated in Feynman's gauge, are obtained in the presence 
of an infrared cutoff, after the removal of an adiabatic switching and with the aid of a suitable mass 
renormalization.\\
In the presence of a dipole approximation, spurious contributions to the infrared factors are shown 
to necessarily arise in Feynman's gauge, with respect both to the Coulomb-gauge model and to the 
amplitudes of Quantum Electrodynamics, and the connection of this result with a recent work on 
the Gupta-Bleuler formulation of non-relativistic models is discussed.\\
It is finally proven that by dropping the dipole approximation and adopting an expansion around a 
fixed charged particle four-momentum, first introduced and employed in the study of the 
infrared problem by Bloch and Nordsieck, the infrared diagrammatic is fully reproduced 
and spurious low-energy effects are avoided.

\ssk\ni 
Key words: quantum electrodynamics; infrared problem; local and covariant gauge; indefinite metric; solvable models

\footnotetext {*email address:simone.zerella@gmail.com}
\thispagestyle {empty}

\newpage

\section* {Introduction}

In Quantum Electrodynamics $(\, QED\, ),\, $ the description of states at asymptotic times and 
the derivation of the scattering matrix are still open issues; it is customary to refer to such 
questions as the infrared ($\, IR\, $) problem.

At the perturbative level, transition amplitudes between states containing a finite number of photons are 
ill-defined, since radiative corrections due to soft photons typically exhibit logarithmic divergences \cite 
{Wei95,JR76}. 
As a consequence, in contrast with ordinary quantum field theories, Dyson's $\, S\: $- matrix \cite {Dys49,Dys51} 
is defined only in the presence of a low-energy cutoff and the problem of a proper identification of asymptotic 
states arises.

As early as 1937, in their pioneering paper on the subject \cite {BN37}, Bloch and Nordsieck proved that 
$IR$ singularities arise in the perturbative expansion because of basic physical facts; they argued that, 
on the basis of the correspondence principle, one has to expect a vanishing probability for the emission of 
a finite number of photons in any collision process involving electrically charged particles. 

Exponentiation of the low-energy photon contributions was conjectured by Schwinger \cite {Sch49} 
and then proved by Yennie, Frautschi and Suura $(YFS)$ in \cite {YFS61} within the framework of the local 
and covariant Gupta-Bleuler formulation \cite {Gup50,Ble50} of $\, QED\, .$
This led to a pragmatic approach to circumvent the soft-photon divergences; one introduces an $\, IR\, $ 
cutoff, sums the transition rates over all final photon states with energy below the threshold of the 
photon detectors and removes the regularization in the resulting expression.
The finiteness of the so-obtained inclusive cross-sections is ensured by cancellations, at each order in 
perturbation theory, between the virtual $\, IR\, $ divergences and those due to soft-photon emission
\cite {Wei95,JR76}.

It is important to remark that such a recipe somehow avoids to take into account the properties 
arising in the characterization of physical charged states, such as the spontaneous breaking of the 
Lorentz boosts in the charged superselection sectors and the absence of a sharp eigenvalue for 
the mass operator of an electrically charged particle.
These features were established in past decades through several model-independent investigations 
\cite {SW74,FPS74,FPS77,FMS79a,FMS79b,Buc82,Buc86}, mostly within the algebraic 
approach to quantum field theory and independently of the perturbative-theoretic framework.

Quite generally, since its early days perturbation theory has been mostly applied to extract physical predictions 
(notable exceptions are \cite {Sym71,MS83,Ste00}) and its relation with the above-mentioned structural 
(non-perturbative) properties of 
$\, QED\, $ is still unclear; in particular, local and covariant quantizations are incompatible with 
positivity \cite {St67} and are only consistent with a generalization of the Wightman axioms.

The above discussion provides motivations to understand and possibly fill the gap that separates the 
Feynman-Dyson formulation from a collision theory in which the structural aspects of the infrared 
problem are taken into account; such an issue is important both conceptually and practically, 
since perturbation theory remains the only source of detailed information on $\, QED\, $ and 
its local and covariant version is the best controlled one regarding the renormalization 
procedure.

As a starting point to address this issue, in the present paper we seek for a mathematical formalization of 
the hypotheses underlying the local and covariant treatment of the $IR$ divergences occurring in 
$\, QED\, .\, $
In particular, we compare the infrared amplitudes of the standard analysis with those obtained from solvable 
hamiltonian models based on two different approximations, which might seem to be equally suited, from a 
physical point of view, for an investigation of low-energy photon effects; the electric dipole approximation 
and the expansion around a fixed (asymptotic) charged particle four-momentum.

Non-relativistic Coulomb-gauge models with a dipole approximation have been studied rigorously in 
\cite {Bla69,Ara83} in order to obtain non-perturbative constructions, in particular of asymptotic  
electromagnetic fields and of one-particle charged states. 
More recently \cite {HS09}, asymptotic e.m. fields and a scattering operator have been constructed 
within the Gupta-Bleuler formulation of non-relativistic $\, QED\, ,$ still in the presence of a dipole 
approximation, by addressing the questions connected with the existence and uniqueness of the 
dynamics of Heisenberg operators in the presence of an indefinite 
metric.

The problem of the comparison of such models with the perturbative structures and 
results has however not been addressed explicitly, thus leaving the validity of perturbative 
procedures in a dubious condition.
In addition, as we shall see, the Gupta-Bleuler formulation of non-relativistic models with a dipole 
approximation suffers from substantial limitations in the comparison with the Feynman-Dyson
expansion.

Our main result \cite {Zer09} is that (only) within hamiltonian models of the ``four-vector Bloch-Nordsieck
 type'' the infrared diagrammatic of the local and covariant expansion of $\, QED\, $ is reproduced by 
M\"{o}ller operators, obtained as weak limits in the presence of a low-energy regularization and a 
suitable mass counterterm, after the removal of an adiabatic switching.
In contrast, the dipole approximation prevents to recover fundamental features of the infrared contributions 
to Feynman's amplitudes; basically, its effects on current conservation, also recognized and studied in 
\cite {HS09}, give rise to substantial additional contributions to the M\"{o}ller operators, 
resulting in particular in a discrepancy by a factor $\, 3/2\, $ with respect to the Coulomb-gauge 
result (and to covariant perturbation theory in $\, QED\, $), in the exponential factors describing 
soft-photon effects.

The structure of the paper is as follows.
Section \ref {sect:1} is devoted to the analysis of the Pauli-Fierz model \cite {FP38}, a non-relativistic 
model formulated in the Coulomb gauge and based on the dipole approximation, taking Blanchard's 
treatment \cite {Bla69} as a starting point.
We prove the existence and uniqueness of the dynamics and obtain M\"{o}ller operators as strong limits 
of the corresponding evolution operator in the interaction representation, for a fixed value of an infrared 
cutoff. 

In order to set up a comparison with the perturbative procedures and results we introduce a four-vector 
model, retaining the approximations of the Pauli-Fierz Hamiltonian. 
Within a simpler setting with respect to \cite {HS09}, we define the dynamics of the model and 
prove its uniqueness; then we construct M\"{o}ller operators as weak asymptotic limits of the 
evolution operator (in the interaction representation) for a fixed infrared cutoff, and discuss 
the spurious effects induced by the dipole approximation.

In Section \ref {sect:2} we introduce models based on an expansion already implicit in \cite {BN37}, 
hereafter referred to as Bloch-Nordsieck $\, (BN)\, $ models.
We will first consider a model formulated in the Coulomb gauge and then its four-vector version.
The existence and uniqueness of the dynamics of the models and the control of the asymptotic limits of the 
evolution operators are achieved within the same framework of Section \ref {sect:1}. 
We show that the infrared amplitudes of $\, QED\, $ are reproduced by means of M\"{o}ller operators
and that the contributions arising in the presence of the dipole approximation disappear.

We conclude the paper with an outlook for future research.

\subsection* {Notations}

\ni
The metric $\, g^{\; \mu\, \nu}=diag\,\, (\, 1\, ,\, -\, 1\, ,\, -\, 1\, ,\, -\, 1\, )$ of Minkowski space 
is adopted and natural units are used $(\, \hbar\, =\, c\, =\, 1\, )\, .\, $
A four-vector is indicated with $\, v^{\, \mu}\, $ or simply with $\, v\, ,\, $ while the symbol $\, \mathbf {v}\, $ 
denotes a three-vector. 
We use the symbol $\, c \cdot d\, $ for the indefinite inner product between four-vectors $\, c\, $ 
and $\, d\, . $ \\   
The norm of a vector $\, \phi\in L^{\, 2}\, $ is indicated by $\, \Vert\, \phi\, \Vert_{\, 2}^{}\, ,$ the Hilbert 
scalar product by $\, (\, .\, ,.\, )\, $ and indefinite inner products by $\, \langle\, .\, ,.\, \rangle\, .\, $\\
The Hilbert-space adjoint of an operator $\, A\, $ is denoted by $\, A^{\; *}, $ while the symbol 
$\, B^{\; \dagger}\, $ stands for the hermitian conjugate, with respect to the indefinite inner 
product $\, \langle\, .\, ,.\, \rangle\, ,\, $ of an operator $\, B\, $ defined on an indefinite-metric 
space.\\
We denote by $\, \mathscr {F}\, $ the symmetric Fock space, by $\, \Vert\, .\, \Vert\, $ the norm 
of $\, \mathscr {F}\, $ and by $\, N\, $ the number operator. 
The projection of $\, \phi\in\mathscr {F}\, $ onto the $\, n\: $- particle space is denoted 
by $\, \phi^{\, (n)}\equiv S_{\; n}^{}\,\, \phi\: ,$ where the orthogonal projection
$\, S_{\; n\, }^{}$ is the symmetrization operator defined in terms of the 
permutation group of degree $\, n\, .\, $\\ 
In the Coulomb-gauge formulation, $\, a_{\; s\; }^{}(\, a_{\; s}^{\; *}\, )$ will stand for the photon annihilation 
(creation) operator-valued distribution, fulfilling the canonical commutation relations $\, (\, CCR\, )\, $
\begin {equation}
\quad [\,\, a_{\; s\; }^{}(\, \mathbf {k}\, )\; ,\; a_{\; s\, '\; }^{\; *}(\, \mathbf {k}\: '\, )\, ]
\; =\,\, \delta_{\: s\, s\, '}^{}\,\,\, \delta\, (\, \mathbf {k}-\mathbf {k}\: '\, )\; ,
\nonumber
\end {equation}
with $\, s\, $ and $\, s\; '\, $ polarization indices.\\
In the same gauge, the Hamiltonian of the free e.m. field is denoted by $\, H_{\: 0\, ,\: tr}^{\; e.\, m.}\, $ and 
the vector potential at time $\, t=0\, $ by
\begin {equation}  
\mathbf {A}_{\, tr\, }^{}(\, \mathbf {x}\, )\, \equiv\; \sum_{s\; }\;\, \int\,\, \frac {\, d^{\,\, 3\, }k} 
{\sqrt {\; 2\;\, \omega_{\,\, \mathbf {k}}^{}}}\;\,\, \mathbf {\epsilon}_{\; s\; }^{}(\, \mathbf {k}\, )
\;\, [\;\, a_{\; s\; }^{}(\, \mathbf {k}\, )\,\,\, e^{\; i\;\, \mathbf {k}\, \cdot\; \mathbf {x}\;\, }+
\, a_{\; s\; }^{\; *}(\, \mathbf {k}\, )\,\,\, e^{\, -\, i\;\, \mathbf {k}\, \cdot\; \mathbf {x}}\;\, ]\,\, ,
\nonumber
\end {equation} 
where $\, \mathbf {\epsilon}_{\; s\, }^{},\; s=1,\, 2,\, $ are orthonormal 
vectors, satisfying the transversality condition 
$\, \mathbf {k}\, \cdot\, \mathbf {\epsilon}_{\; s\,\, }^{}
(\, \mathbf {k}\, )=\, 0\; .\, $\\
The annihilation and creation operator-valued distributions in the $\, FGB\, $ gauge, denoted respectively 
by $\, a^{\, \mu\, }(\, \mathbf {k}\, )\, $ and $\, a^{\; \mu\; \dagger}\, (\, \mathbf {k}\, )\, ,\, $ fulfill the 
$\, CCR\, $ 
\begin {equation}
[\,\, a^{\, \mu}\, (\, \mathbf {k}\, )\: ,\; a^{\: \nu\; \dagger\, }(\, \mathbf {k}\, '\, )\; ]\, =
\, -\; g^{\: \mu\, \nu}\,\,\, \delta\; (\, \mathbf {k}-\mathbf {k}\; '\, )\; .
\nonumber
\end {equation}
In the same gauge, the Hamiltonian of the free e.m. field is denoted by $\, H_{\; 0}^{\,\, e.\, m.\, }\, $ and 
the vector potential at time $\, t\, =\, 0\, $ by
\begin {equation}  
A^{\; \mu\; }(\, \mathbf {x}\, )\, \equiv\, \int\,\, \frac {\, d^{\,\, 3\, }k} 
{\sqrt {\; 2\;\, \omega_{\,\, \mathbf {k}}^{}}}\;\;\, 
[\,\, a^{\, \mu}\, (\, \mathbf {k}\, )\;\,\, 
e^{\; i\,\, \mathbf {k}\, \cdot\; \mathbf {x}}\, +\, 
a^{\, \mu\: \dagger}\, (\, \mathbf {k}\, )
\;\,\, e^{\, -\, i\;\, \mathbf {k}\, \cdot\; \mathbf {x}}\;\, ]\; .
\nonumber
\end {equation} 
We write
\begin {equation}  
F^{\; \mu\, \nu}\, (\, \mathbf {x}\, )\, \equiv\;\, \partial^{\; \mu}\,\, A^{\; \nu\; }(\, \mathbf {x}\, )-\,
\partial^{\; \nu}\,\, A^{\; \mu\; }(\, \mathbf {x}\, )
\quad\quad\nonumber
\end {equation}
for the e.m field tensor at $\, t\, =\, 0\, .$\\
The convolution with a form factor $\, \rho\, $ is indicated by
\begin {equation}\label {convoluzione covariante}
A^{\; \mu}\; (\, \rho\, ,\; \mathbf {x}\; )\, \equiv\; \int\,\, d^{\,\, 3\, }\xi\;\,\,\, \rho\,\, (\, \mathbf {\xi}\, )
\;\,\, A^{\; \mu}\; (\, \mathbf {x}-\, \mathbf {\xi}\; )\; ,
\nonumber
\end {equation}
and similarly for $\, \mathbf {A}_{\, tr}^{}\, .\, $
For brevity we write
\begin {equation}
a\, (f\, (\, t\, ))\, \equiv\; \int\,\, d^{\,\, 3\, }k\;\,\, a^{\: \mu\; }(\, \mathbf {k}\, )
\,\,\, f_{\, \mu}^{}\: (\, \mathbf {k}\, ,\; t\, )\;\,\,
\nonumber
\end {equation}
and denote the corresponding sum in the Coulomb gauge by $\, a_{\, tr\, }^{}(f\, (\, t\, ))\, .$\\ 
$\, \mathscr {S}\, (\, \mathbb {R}^{\, 3}\, )\, $ will stand for the Schwartz space of 
$\, C^{\; \infty}\, $ functions of rapid decrease on $\, \mathbb {R}^{\; 3\; }.$

\section {Pauli-Fierz-Blanchard Models}
\label {sect:1}

In the present Section we discuss the Pauli-Fierz-Blanchard $(PFB)$ model and formulate a 
suitable four-vector model, retaining the approximation of the Pauli-Fierz hamiltonian.

The model \cite {FP38} describes the interaction of a spinless Schr\"{o}dinger particle with the 
quantum electromagnetic field, under suitable infrared approximations. 
It was reconsidered three decades later by Blanchard, who investigated the questions connected with 
a mathematical formulation of the fact that an infinite number of photons is emitted in any collision 
process involving electrically charged particles.
In \cite {Bla69}, he proved the existence of the dynamics and showed that a unitary operator 
can be obtained as the limit of evolution operators \emph {in the sense of morphisms of a suitable 
($\, C^{\, *}$-) algebra}.
Furthermore, he established the existence of M\"{o}ller operators \emph {interpolating between the 
Pauli-Fierz Hamiltonian and its perturbation by a potential term}, for a large class of potentials.

The model was further studied by Arai \cite {Ara83}, who was able to construct one charged-particle 
states and obtained a scattering operator at fixed charged particle momentum, also allowing for the 
inclusion in the hamiltonian of the bilinear term in the gauge field. 
In the sequel we shall take Blanchard's treatment as a starting point, although our analysis will
require some changes with respect to his setting.

First, since we wish to employ the model in order to investigate the methods at the basis of the local and 
covariant perturbative treatment of the infrared divergences, we adopt an $\, IR\, $ cutoff throughout
the analysis. 
Even in its presence, the limits considered by Blanchard for the M\"{o}ller operators only exist in the weak 
topology and do not define unitary operators; in order to obtain strong convergence and unitarity, a mass 
renormalization and an adiabatic switching of the interaction will be essential.

Second, we introduce a four-vector version of the model, prove the existence and uniqueness of the
dynamics and control the large-time limits of the time-evolution operator in the interaction representation, 
yielding M\"{o}ller operators as isometries on an indefinite vector space.

We consider the infrared-regularized $\, PFB\, $ Hamiltonian
\begin {equation}\label {hamiltoniano PFB}  
H_{\: \lambda}^{\,\, (PFB)}\, =\, \frac {\,\,\, {\mathbf {p}}^{\, 2}} {2\; m}\, +
\, H_{\; 0\, ,\,\, tr\; }^{\,\, e.\, m.}+\, H_{\, int\, ,\; tr\, }^{}\equiv\; H_{\; 0\, }^{}+\, H_{\, int\, ,\; tr}^{}\,\, ,
\end {equation}
\begin {equation}\label {interazione PFB}
H_{\, int\, ,\; tr}^{}=\, -\, \frac {e} {m\, }\;\;\, \mathbf {p}\, \cdot
\mathbf {\, A}_{\, tr}^{}\, (\, \rho\: ,\, \mathbf {x}= 0\; )\; .
\end {equation}
The particle, of mass $\, m\, ,\, $ charge $\, e\, $ and rotationally invariant distribution of charge 
$\rho\in\mathscr {S\, }(\, \mathbb {R}^{\, 3}\, )\, ,$ will also be called electron. 
The subscript $\, \lambda\, $ on the left-hand side of (\ref {hamiltoniano PFB}) denotes the fictitious 
photon mass employed as an infrared-regularization method as in $\, QED\, ,$ by setting 
$\, \omega_{\; \mathbf {k}}^{\; 2}\equiv\mathbf {k}^{\; 2}+\, \lambda^{\; 2},\, $ and 
the $\, \lambda\; $- dependence in $H_{\; 0\, ,\,\, tr\, }^{\; e.\, m.}\, $ and in 
(\ref {interazione PFB}) is understood.
The functional form of the interaction is dictated by the electric dipole approximation and implies that the 
electron momentum is conserved, while the total one is not.

The Hilbert space of states of the model is $\, \mathscr {H}=\, L^{\; 2\,\, }(\, \mathbb {R}^{\, 3}\, )\otimes
\mathscr {F\, },$ with $\, L^{\; 2}\, $ the one-particle space and $\, \mathscr {F}\, $ the Fock space of 
photons.
Since $\, H_{\; 0}^{}\, $ is the sum of two positive and commuting self-adjoint operators, 
it is self-adjoint; in particular, it is essentially self-adjoint (e.s.a.) on 
$\, D_{\; 0}^{}=\mathscr {S\, }(\, \mathbb {R}^{\, 3}\, )\otimes D_{\, F_{\: 0\, }^{}}^{},$ 
with 
$\, D_{\, F_{\: 0}^{}}^{}\equiv\, (\: \psi\in F_{\; 0\; }^{};\, \psi^{\, (n)}\in S_{\; n}^{}\, 
\bigotimes_{\; k\; =\,\, 1\; }^{\; n}\mathscr {S\, }(\, \mathbb {R}^{\, 3}\, )\, ,\,
\forall\; n\, )\, ,\, F_{\,\, 0}^{}$ the set of finite particle vectors of 
$\, \mathscr {F}\, .$

Next we prove the essential self-adjointness of the Hamiltonian (\ref {hamiltoniano PFB}), making use of
techniques already exploited in \cite {Nel64, Ara81}.\\
By the estimate $\, \Vert\; b_{\; s}^{}\, (\, f\, )\; \Psi\, \Vert\leq\Vert\, f\, \Vert_{\; 2}^{}\; \Vert\, (\, N+\, \mathbb {I}\, )^{\; 1/\, 2}\; \Psi\, \Vert\, ,\, $ $\forall\; \Psi\in Dom\, (\, N^{\; 1/\, 2}\, )\, ,$ where $b_{\; s}^{}\, (\, f\, )\, ,$ $\, f\in L^{\; 2}\, ,\, $ stands for either $\, a_{\; s\; }^{}(\, f\, )\, $ or $\, a_{\; s\; }^{\; *}(\, f\, )\, ,\, $ one finds
\begin {equation}
\Vert\, \mathbf {A}_{\: tr}^{\, i}\, (\, \rho\, ,\, \mathbf {x}= 0\; )\,\, \Psi\, \Vert\, \leq\, c\, (\, \rho\, )\,\, 
\Vert\, (\, N +\, \mathbb {I}\, )^{\; 1\, /\, 2}\;\, \Psi\, \Vert\; ,\; \forall\; \Psi\in D_{\, F_{\: 0\, }^{}}^{},\, 
\end {equation}
with $\, c\, (\, \rho\, )\, $ is a (positive) constant, provided the form factor $\, \rho\, $ is held fixed.
Further, for any $\, \lambda > 0\, ,\, \Vert\, (\, N +\, \mathbb {I}\, )\,\, \Psi\, \Vert\leq\lambda^{\; -\, 1}
\,\, \Vert\, H_{\; 0\, ,\; tr\; }^{\,\, e.\, m.}\: \Psi\, \Vert +\Vert\, \Psi\, \Vert\, ,\, \forall\, \Psi\in 
D_{\, F_{\: 0\, }^{}}^{},$ hence, by virtue of the estimate 
$$ 2\;\, \Vert\, (\, A\, \otimes\, B\, )\; \Phi\, \Vert\leq \Vert\, (\, A^{\; 2}\otimes\, \mathbb {I}\, +
\, \mathbb {I}\, \otimes\, B^{\; 2}\, )\; \Phi\, \Vert\, ,\; \forall\; \Phi\in D_{\; 0}^{}\, , $$
with $\, A=\mathbf {p}\, ,\, B=(\, N+\, \mathbb {I}\, )^{\; 1\, /\: 2}\, ,$  one gets the 
bound
\begin {equation}\label {bound autoaggiunzione}
\Vert\, H_{\: \lambda}^{\,\, (PFB)}\,\, \Phi\, \Vert\, \leq
\, d\; (\, e\, ,\, \lambda\, ;\, \rho\, )\,\, \Vert\, (\, H_{\; 0}^{}+\, \mathbb {I}\, )
\,\, \Phi\, \Vert\: ,\; \forall\; \Phi\in D_{\; 0}^{}\, ,\, 
\end {equation}
for a suitable function $d\, (\, e\, ,\lambda\, ;\, \rho\, )\, . $
By the bound (\ref {bound autoaggiunzione}) and the $CCR$, there exists 
$g\, (\, e\, ,\, \lambda\, ;\, \rho\, )$ such that $\vert (\, \Phi\, ,\, [\, H_{\, \lambda}^{\; (PFB)},
H_{\; 0\; }^{}]\: \Phi\, )\vert\leq g\, (\, e\, ,\, \lambda\, ;\, \rho\, )\, \Vert(\, H_{\; 0}^{}+
\, \mathbb {I}\, )^{\: 1/\, 2\,\, }\Phi\, \Vert\, ,$ $\forall\: \Phi\in D_{\; 0\; }^{};$ Nelson's 
commutator theorem, in the formulation given by Faris and Lavine 
\cite {FL74,RS72a}, then implies that $\, H_{\: \lambda}^{\; (PFB)}\, $ 
is e.s.a. on $\, D_{\; 0}^{}\, ,\, $ for all $\, \vert\: e\: \vert\, ,\, 
\lambda >0\, $.\\
Since the Hamiltonian commutes with the electron momentum operator, it is e.s.a.\footnote {The e.s.a. 
of the hamiltonian at fixed \textbf {p} also follows by the Kato-Rellich theorem.} on (almost) any of the 
subspaces, on which $\, \mathbf {p}\, $ takes a constant value, obtained by decomposing $\, D_{\; 0}^{}\, $ 
on the joint spectrum of the components of $\, \mathbf {p}\, $; without losing generality we consider 
therefore $\, \mathbf {p}\, $ fixed in the rest of the analysis of the $\, PFB\, $ model.

The time-evolution in the interaction picture, corresponding to an Hamiltonian $\, H =H_{\; 0}^{}+
H_{\, int}^{}\, $, with $\, H_{\; 0}^{}\, $ the free part and $\, H_{\, int}^{}\, $ a time-independent 
interaction, is governed by the operator
\begin {equation}\label {UI}
U_{\,\, I}^{}\; (\, t\, )\, \equiv\;\, \exp\; (\; i\; H_{\; 0}^{}\,\, t\; )\,\,\, \exp\; (\, -\, i\,\, H\,\, t\; )\, 
\end {equation}
and satisfies, on $\, D_{\; 0}^{}\, ,$
\begin {equation}\label {equazione di evoluzione interazione}
i\,\,\, \frac {\, d\,\, U_{\,\, I}^{}\; (\, t\, )} {d\,\, t}\, =\; H_{\; I}^{}\; (\, t\, )\;\,\, U_{\,\, I}^{}\; (\, t\, )\; ,
\end {equation}
with
$$
H_{\; I\;\, }^{}(\, t\, )\, \equiv\;\, \exp\; (\; i\; H_{\; 0}^{}\,\, t\; )\,\,\, H_{\, int}^{}\;\, \exp\; (\, -\, i\; H_{\: 0}^{}\,\, t\; )\, .
$$ 
(\ref{equazione di evoluzione interazione}) identifies $\, U_{\; I\; }^{}(\, t\, )\, $ (a more general result, 
also applying to time-dependent interactions, is stated below).
Since the commutator of $ H_{\; I\, }^{}$ evaluated at different times is a multiple of the identity 
operator at each definite momentum, we make use of the formula \cite {Wil67}
\begin {equation}\label {caso particolare formula BH}
\quad\quad\quad e^{\: A\, +\, B}\, =\; e^{\, A}\;\,\, e^{\, B}\;\,\, e^{\, -\frac {1} {2}\;\, [\, A\, ,\; B\; ]}\; .
\end {equation}
Here and in the following, (\ref{caso particolare formula BH}) will be used to compute evolution operators, 
with an a posteriori verification that the so-obtained operators satisfy the evolution equation on a suitable 
domain, here on a core of the common domain of the operators $\, H_{\; I}^{}\, (\, t\, )\, .\, $  
We obtain
\begin {eqnarray}\label {operatore di evoluzione a tempi finiti}
U_{\,\, I\,\,\, }^{}(\, t\, )\, =\,\, \exp\,\, (\, -\,\, i\; \int_{\, 0}^{\; t\; }d\,\, t\; '
\,\,\, H_{\,\, I\,\, }^{}(\, t\; '\, )\, )\;\; \exp\,\, (\, -\, \frac {1} {2}\,\, 
\int_{\, 0}^{\; t}\; d\,\, t\; '
\nonumber\\
\quad\quad\quad\quad\quad\quad\quad\quad\quad\quad\times\, \int_{\, 0}^{\; t\; '} d\,\, t\; ''\,\,\, 
[\,\, H_{\,\, I\;\, }^{}(\, t\; '\, )\; ,
\, H_{\,\, I\;\, }^{}(\, t\; ''\, )\; ]\, )\; ,
\end {eqnarray}
hence $\, U\: (\, t\, )\, =\,\, \exp\,\, (\, -\, i\,\, H_{\; 0}^{}\;\, t\; )\;\,\, U_{\,\, I\;\, }^{}(\, t\, )\, .\, $
By explicit calculations one gets
\begin {equation}\label {operatore di evoluzione non rin}
U_{\,\, I\;\, }^{}(\, t\, )\, =\;\, c\; (\, t\, )\,\,\, \exp\,\, (\,\, i\; e\,\, (\; a_{\; tr\; }^{\; *}(\, f_{\; \mathbf {p}}^{}\; (\, t\, ))\, +
\, a_{\; tr\; }^{}(\, \overline {f\, }_{\mathbf {p}}^{}\; (\, t\, ))\, )\, )\: ,
\quad
\end {equation}
\begin {equation}\label {contributo numerico tempo fisso}
c\,\, (\, t\, )\, =\;\, \exp\,\, (\,\, \frac {\, i\,\, e^{\, 2}\,\, {\mathbf {p}}^{\, 2}}
{3\,\, m^{\, 2}}\,\, \int\; \frac {\, d^{\,\, 3\, }k} {\omega_{\; \mathbf {k}\, }^{\,\, 2}}\;\, {\tilde {\rho\; }^{2\,\, }
(\, \mathbf {k}\, )}\,\,\, (\; t\, -\frac {\, \sin\; \omega_{\; \mathbf {k}}^{}\,\, t} {\omega_{\; \mathbf {k}}^{}}
\,\, )\, )\: ,
\end {equation}
\begin {equation}\label {funzione di coerenza tempo fisso}
f_{\; \mathbf {p}\, s\,\, }^{}(\, \mathbf {k}\, ,\: t\; )\, =
\; \frac {\; \tilde {\rho\; }(\, \mathbf {k}\, )} 
{\sqrt {\; 2\;\, \omega_{\,\, \mathbf {k}}^{}}\, }
\;\, \frac {\; \mathbf {p}\, \cdot\, 
\mathbf {\epsilon}_{\; s\; }^{}(\, \mathbf {k}\, )} {m\; }\;\, 
\frac {\, e^{\,\, i\;\, \omega_{\; \mathbf {k}}^{}\; t}-\, 1\, }
{i\;\, \omega_{\,\, \mathbf {k}}^{}}\; \cdot\;\;
\end {equation}
It is easy to check that $\, U_{\; I\;\, }^{}(\, t\, )\, $ does not converge for large times; 
thus we introduce a mass counterterm and a regularization of the oscillating terms occurring 
in (\ref {contributo numerico tempo fisso}), (\ref {funzione di coerenza tempo fisso}), by 
replacing the electric coupling by $\, e^{\: (\, ad\, )}\, (\, t\, )\equiv\, e\;\, 
e^{\, -\, \epsilon\,\, \vert\, t\, \vert\; }.\, $
The resulting Hamiltonian is 
\begin {equation}\label {hamiltoniano PFB rinormalizzato}
H_{\: \lambda\; ,\; R}^{\; (PFB)}\, =\, \frac {\;\, \mathbf {p}^{\; 2}} {2\,\, m\, }+
\, H_{\; 0\, ,\,\, tr\, }^{\,\, e.\, m.\, }+\, H_{\, int\, ,\; tr\, ,\; R}^{\; (\, \epsilon\, )}\;\, ,
\quad\quad\quad\quad\quad\quad
\end {equation}
\begin {equation}\label {interazione PFB rinormalizzata}
H_{\, int\, ,\; tr\, ,\; R}^{\; (\, \epsilon\, )}\, \equiv\; H_{\, int\, ,\; tr}^{}\;\,\, e^{\, -\, \epsilon\,\, \vert\, t\, \vert\, }+
\, z\;\, e^{\, 2}\;\, \frac {\;\,\, \mathbf {p}^{\, 2}} {2\,\, m}\;\;\, 
e^{\, -\, 2\:\, \epsilon\; \vert\, t\, \vert}\,\, ,
\end {equation}
\begin {equation}\label {controtermine PFB}
z\; =\, \frac {2} {3\,\, m\, }\,\, \int\,\, \frac {\, d^{\,\, 3\, }k} {\omega_{\,\, \mathbf {k}}^{\,\, 2}\, }
\;\;\, \tilde {\rho\; }^{2\; }(\, \mathbf {k}\, )\, .\quad\quad\quad\quad
\end {equation}
In the sequel, we state the results for positive times; in order to obtain the corresponding expressions for 
$\, t<0\, ,\, $ it suffices to replace $\, \epsilon\, $ by $\, -\, \epsilon\, $.
Although the above Hamiltonian is time dependent, existence and uniqueness of the time-evolution unitary 
operators follow from the independence of time of their selfadjointness domain and strong differentiability 
of $\, (\, H\, (\, t\, )-\, i\, )\, (\, H\, (\, 0\, )-\, i\, )^{\, -\, 1}\, ,\, $ through the results of \cite {Kat56}.\\ 
Upon inserting (\ref {interazione PFB rinormalizzata}), (\ref {controtermine PFB}) into the right-hand 
side (r.h.s.) of  (\ref {operatore di evoluzione a tempi finiti}) we obtain
\begin {equation}\label {operatore di evoluzione PF rinormalizzato}
U_{\; I\, ,\,\, tr\, ,\; \lambda\,\, }^{\,\, (\, \epsilon\, )}(\, t\, )\,  \equiv\;\, c_{\,\, z}^{\,\, (\, \epsilon\, )}\, (\, t\, )
\,\,\, \exp\,\, (\,\, i\; e\;\, (\; a_{\; tr}^{\; *}\, (\, f_{\; \mathbf {p}}^{\,\, (\, \epsilon\, )\; }(\, t\, ))\, 
+\, a_{\; tr}^{}\, (\, \overline {f\, }_{\mathbf {p}}^{\,\, (\, \epsilon\, )\; }(\, t\, ))\, )\, )\: ,
\end {equation}
with
\begin {equation}\label {contributo numerico tempo fisso regolarizzato}
c_{\,\, z}^{\,\, (\, \epsilon\, )}\, (\, t\, )\, \equiv\;\, \exp\,\, (\,\, \frac {\, i\; e^{\, 2}\,\, 
\mathbf {p}^{\, 2}} {3\; m^{\, 2}}\;\,\, d^{\,\, (\, \epsilon\, )}\, (\, t\, )\, )
\,\,\, \exp\,\, (\; i\; e^{\, 2}\; z\;\, \frac {\;\,\, \mathbf {p}^{\: 2}} 
{2\; m\, }\;\, \frac {\, e^{\, -\, 2\,\, \epsilon\:\, t}\, -\, 1\, } 
{2\,\, \epsilon\, }\; )\: ,
\end {equation}
\begin {equation}\label {contributo numerico esplicito}
\quad d^{\,\, (\, \epsilon\, )}\; (\, t\, )\, =\, -\int\,\, \frac {d^{\,\, 3\, }k\;\;\, {\tilde {\rho\; }}^{2\,\, } 
(\, \mathbf {k}\, )} {\omega_{\; \mathbf {k}}^{}\, (\, \omega_{\; \mathbf {k}}^{\,\, 2\, }+\, 
\epsilon^{\; 2}\, )}\,\,\, (\,\, e^{\, -\, \epsilon\,\, t}\;\, \sin\; \omega_{\; \mathbf {k}}^{}
\,\, t\; +\, \frac {\, \omega_{\; \mathbf {k}}^{}} {2\,\, \epsilon}\,\,\, 
(\,\, e^{\, -\, 2\,\, \epsilon\,\, t\, }-\, 1\; )\, )\; ,
\end {equation}
\begin {equation}\label {funzione di coerenza regolarizzata}
f_{\; \mathbf {p}\, s\;\,\, }^{\: (\, \epsilon\, )}(\, \mathbf {k}\, ,\; t\, )\, =\, \frac {\; \tilde {\rho\; }(\, \mathbf {k}\, )} 
{\sqrt {\; 2\;\, \omega_{\; \mathbf {k}}^{}}\, }\;\, \frac {\; \mathbf {p}\, \cdot\, \mathbf {\epsilon}_{\; s\; }^{}
(\, \mathbf {k}\, )} {m\; }\;\, \frac {\, e^{\; (\: i\,\, \omega_{\: \mathbf {k}\, }^{}-\; \epsilon\, )\;\, t}\, -\, 1\, }
{i\;\, \omega_{\; \mathbf {k}}^{}-\, \epsilon}\, \cdot
\end {equation}
(\ref {contributo numerico tempo fisso regolarizzato}),(\ref {contributo numerico esplicito}) provide a 
regularization of (\ref {contributo numerico tempo fisso}); in particular, the oscillating term on the r.h.s. 
of (\ref {contributo numerico esplicito}) vanishes for $\, t\, \rightarrow\infty\, ,$ due to the presence of 
the adiabatic factor, and the residual contribution of order $\, 1\, /\, \epsilon\, $ from 
$\, d^{\,\, (\, \epsilon\, )}\, (\, t\, )\, $ 
to the first exponential on the r.h.s of (\ref {contributo numerico tempo fisso regolarizzato}) is canceled, 
for $\, \epsilon\rightarrow 0\, $, by the $\, z\; $- dependent exponential.
Hence the existence of the asymptotic time limits and of the adiabatic limit of (\ref {contributo numerico 
tempo fisso regolarizzato}) is proven.

The strong convergence of the evolution operator (\ref {operatore di evoluzione PF rinormalizzato}) can 
now be established:
\begin {theorem}\label {proposizione Moller PFB}
By choosing the coefficient of the mass counterterm as in (\ref {controtermine PFB}), both the large-time 
limits and the adiabatic limit of the evolution operator (\ref {operatore di evoluzione PF rinormalizzato}), 
defining the M\"{o}ller operators, exist in the strong topology of $\, \mathscr {H}:$
\begin {eqnarray}\label {Moller PFB}
\Omega_{\; \pm\, ,\,\, tr}^{}=\;\, s\; -\, \lim_{\epsilon\; \rightarrow\; 0}\; \lim_{t\; \rightarrow\, \mp\, \infty}
\,\,\, U_{\,\, I\, ,\,\, tr\, ,\; \lambda\,\, }^{\,\, (\, \epsilon\, )}(\, -\; t\; )\, =
\quad\quad\quad\quad\quad\quad\quad\quad\quad\quad\nonumber\\
=\;\, \exp\,\, (\, -\; i\; e\;\, \sum_{s\; }\;\;\, [\,\, a_{\; s}^{\; *}\; (\, f_{\; \mathbf {p}\, s}^{}\, )\, +
\, a_{\; s}^{}\: (\, \overline {f\, }_{\mathbf {p}\, s}^{}\, )\; ]\, )\: ,
\end {eqnarray}
\begin {eqnarray}\label {limiti forti}
f_{\; \mathbf {p}\, s\,\, }^{}(\, \mathbf {k}\, )\, \equiv\,\, L^{\; 2\, }-\lim_{\epsilon\,\, \rightarrow\; 0\, }
\; f_{\; \mathbf {p}\, s\,\, }^{\; (\, \epsilon\, )\; }(\, \mathbf {k}\, )\, \equiv\,\, L^{\; 2\, }-
\lim_{\epsilon\; \rightarrow\; 0}\, \lim_{t\, \rightarrow\, \mp\, \infty}\;
f_{\; \mathbf {p}\, s\;\,\, }^{\; (\, \epsilon\, )}(\, \mathbf {k}\, ,\: t\; )\, =
\nonumber\\\quad\quad\quad\quad\quad\quad\;\,  
=\, \frac {\; \tilde {\rho\; }(\, \mathbf {k}\, )} {\sqrt {\; 2\;\, \omega_{\; \mathbf {k}}^{}}\, }
\;\, \frac {\; \mathbf {p}\, \cdot\, \mathbf {\epsilon}_{\; s\; }^{}(\, \mathbf {k}\, )} {m\; }\;\, 
\frac {\, i\, }{\omega_{\; \mathbf {k}}^{}}\, \cdot
\end {eqnarray}
\end {theorem}
\noindent
\begin {proof} 
The limits of $\, f_{\; \mathbf {p}\, s\;\, }^{\; (\, \epsilon\, )}(\, \mathbf {k}\, ,\, t\, )\, $ exist pointwise and 
thus in the strong $\, L^{\; 2}\, $ topology by the dominated convergence theorem.

The operator $\Phi\, (f_{\; \mathbf {p}}^{\; (\, \epsilon\, )}(\, \mathbf {k}\, ,\, t\, ))\equiv 
a_{\; tr\, }^{}(f_{\,\, \mathbf {p}}^{\; (\, \epsilon\, )}(\, \mathbf {k}\, ,\, t\, ))+\, 
a_{\; tr\, }^{\; *}(\, \overline {f}_{\, \mathbf {p}}^{\; (\, \epsilon\, )}
(\, \mathbf {k}\, ,\, t))$ admits $\, F_{\,\, 0}^{}\, $ as a dense 
and invariant set of analytic vectors and is therefore
e.s.a. on $F_{\,\, 0}^{}$ due to Nelson's analytic 
vector theorem \cite {RS72a}.
Linearity and Fock-space estimates imply that $\, \Phi\, (f_{\; \mathbf {p}\;\, }^{\, (\, \epsilon\, )\, }
(\, \mathbf {k}\, ,\, t\, ))$ converges strongly to $\, \Phi\, (f_{\; \mathbf {p}\; }^{\, (\, \epsilon\, )\, }
(\, \mathbf {k}\, ))\, $ on $\, F_{\:\, 0}^{}\, .\, $
Since $\, \Phi\, (f_{\; \mathbf {p}\; }^{\, (\, \epsilon\, )\, }(\, \mathbf {k}\, ))\, $ is e.s.a. on $\, F_{\,\, 0}^{}\, ,$ one 
has convergence in the strong generalized sense and by Trotter's theorem \cite {RS72b} the existence 
of the time-limits in (\ref {Moller PFB}) follows.
A similar proof allows to establish the strong convergence of $\, \Phi\, (f_{\; \mathbf {p}\;\, }^{\, (\, \epsilon\, )\, }
(\, \mathbf {k}\, ))\, $ for $\, \epsilon\rightarrow 0\, .$ 
\end {proof}
\ni
We wish to remark that without the adiabatic regularization the time limits of the operators considered above 
would only exist as weak limits; as a matter of fact, the Riemann-Lebesgue lemma implies the existence of the weak 
limits $\, w-\lim_{\; t\rightarrow\, \mp\, \infty\, }f_{\; \mathbf {p}\, s}^{\; (\, \epsilon\, =\, 0\, )\, }(\, \mathbf {k}\, ,\, t\, )=
f_{\; \mathbf {p\, }s}^{}\, (\, \mathbf {k}\, )\, ,$ while $\, \Vert\, f_{\; \mathbf {p}\, s\,\, }^{\; (\, \epsilon\, =\, 0\, )\, }(\, t\, )-
f_{\; \mathbf {p}\, s\,\, }^{}\Vert_{\; 2}^{}\, $ does not converge to zero.

Treating the model as a description of the soft-photon effects in $\, QED\, $ and introducing a unitary operator
$\, \mathscr {W}\, $ on $ L^{\, 2\,\, }(\, \mathbb {R}^{\, 3}\, )\, $, interpreted as a scattering operator for 
electrons, acting at time zero, a non-trivial $\, S\; $- matrix is given, at fixed $\, \lambda\, ,\, $ by 
\begin {eqnarray}\label {matrice di scattering PFB}
S_{\,\, \lambda}^{}\, \equiv\;\, s\; -\, \lim_{\epsilon\: \rightarrow\: 0}\,\, \lim_{t\, ,\; t\, '\, \rightarrow\, +\infty}
\; U_{\,\, I\, ,\; tr\, ,\; \lambda\,\, }^{\; (\, \epsilon\, )}(\, t\, )\;\,\, \mathscr {W}\;\,\, 
U_{\,\, I\, ,\,\, tr\, ,\; \lambda\,\, }^{\; (\, \epsilon\, )}(\; t\; '\; )\: .
\end {eqnarray}

In order to compare the infrared diagrammatic of $\, QED\, $ with the expansion of M\"{o}ller's operators in 
powers of the electric charge, it is necessary to formulate a model in a gauge employing four independent 
photon degrees of freedom, such as Feynman's gauge.
With this aim, we introduce, as in  \cite {HS09}, a four-vector model retaining the approximations of the 
Pauli-Fierz Hamiltonian.

The model is defined in the tensor product space $\, \mathscr {V}\equiv\, L^{\: 2\,\, }(\, \mathbb {R}^{\, 3}\, ) 
\otimes\, \mathscr {G}\, $, $\, \mathscr {G} $ an indefinite-metric photon space to be constructed below.
The Hamiltonian is given, on the domain $\, \mathscr {V}_{\,\, 0}^{}\, $ defined in (\ref {definizione V_0}), by
\begin {equation}\label {hamiltoniana PFBR} 
\; H_{\: \lambda\, }^{\; (PFBR\, )\, }=\; m\, +\, \frac {1} {2}\,\, m\; {\mathbf {v}}^{\; 2\, }+
\, H_{\; 0}^{\;\, e.\, m.\, }+\, e\;\, \tilde {v\, }\, \cdot\, A\; (\, \rho\, ,\, \mathbf {x}=0\; )\, 
=\, H_{\; 0\, }^{}+\, H_{\, int}^{}\; ,
\end {equation}
with $\, \tilde {v\; }^{\mu}=(\, 1+{\, \mathbf {v}}^{\, 2}/\, 2\, ,\mathbf {v}\, )\, ,
\mathbf {\, v}\equiv\, \mathbf {p}\, /\, {m}\, ,$ and will be referred to as 
$\, PFBR\, $ Hamiltonian.

In general, for models employing covariant vector potentials, non positivity of the scalar product raises 
substantial questions on selfadjointness and existence and uniqueness of time evolution.
In \cite {HS09} such problems have been treated on a slightly different version of the 
model, providing a general framework for existence and uniqueness of the Heisenberg time-evolution within 
a Hilbert-space formulation.

The same methods could be used to discuss the evolution operators as unbounded operators on the Hilbert 
space, preserving the indefinite scalar product on a dense domain.
Since, however, the Hamiltonian (\ref {hamiltoniana PFBR}) is quadratic in the e.m. field variables and only 
involves a commutative algebra in the charged particle variables, a more pragmatic approach is enough for 
our purposes, with time-evolution operators defined in terms of Weyl exponentials of fields, introduced 
starting from their algebraic relations, on a suitable invariant vector space.

Uniqueness of the solution will be obtained by observing that our space can be identified with a dense 
domain of the Hilbert space introduced in \cite {HS09}, that our time-evolution operators, 
although unbounded, are continuous and differentiable on $\, \mathscr {V}_{\;\, 0}^{}$ in the Hilbert 
strong topology and that any group of isometries of $\, \mathscr {V}_{\;\, 0}^{}\, $, differentiable in 
the strong sense and with derivative given by (\ref {hamiltoniana PFBR}), coincides with them. 

The space $\, \mathscr {G} \, $ is defined as follows. Let $\, \mathscr {A}_{\,\, ext}^{\:\, e.\, m.}\, $ be the 
${}^{*}$-algebra generated by the photon canonical variables and by variables (Weyl operators 
in momentum space) $\, W\, (\, g\, ,\, h\, )\, ,\, $ indexed by four-vector real-valued functions in 
$ L^{\, 2\; }(\, \mathbb {R}^{\, 3}\, )\, ,\, $ fulfilling
\begin {equation}\label {proprieta' W}
W\, (\, g\; ,\; h\, )^{\, *}=\; W\, (\, -\, g\: ,-\, h\, )\; ,\; 
W\, (\, g\; ,\; h\, )^{\, *}\;\, W\, (\, g\; ,\; h\, )\, =\, 1\: ,
\end {equation}
\begin {equation}\label {commutatore W}
W\, (\, g\; ,\; h\, )\;\, W\, (\; l\; ,\, m\, )\, =\; \exp\; (\; i\,\, (\langle\, g\; ,\, m\; \rangle-\langle
\, h\: ,\: l\; \rangle)\, )\;\,\, W\, (\; l\; ,\, m\, )\;\, W\, (\, g\; ,\; h\, )\, ,
\end {equation}
\begin {equation}\label {commutatore a W}
[\; a\, (\, \overline {f\, }\, )\, ,\, W\, (\, g\; ,\; h\, )\, ]\, =\frac {\, i} {\sqrt{\, 2\, }}\;\, 
\langle\, f\, ,\: n\; \rangle\,\,\, W\, (\, g\; ,\; h\, )\; ,\; n\, \equiv\,\, g\, +\, i\; h\; .
\end {equation}
In the above formulae the symbol $\, {}^{*}\, $ stands for the algebra involution and
\begin {equation}\label {prodotto indefinito}
\langle\, f\, ,\, g\; \rangle\, \equiv\,\, (\, f^{\; 0\, },\, g^{\; 0}\, )\, -\, 
\sum_{i\, }\,\,\, (\, f^{\; i\, },\, g^{\; i}\, )\: .
\end {equation}
The Fock functional is characterized by the following expectations over $\, \mathscr {A}_{\,\, ext}^{\,\, e.\, m.}$
\begin {equation} 
\label{propagatore Fock}
\omega_{\; F}^{}\; (\, a\, (\, \overline {f\, }_{1}\, )\,\,\, a^{\, *}\, (\, f_{\; 2}^{}\, ))\, =
-\, \langle\, f_{\; 1}^{}\: ,\, f_{\; 2}^{}\; \rangle\: ,
\quad\quad\quad\quad\quad\quad\quad
\end {equation}
\begin {equation}\label {aspettazione esponenziale}
\omega_{\; F}^{}\; (\, W\, (\, g\: ,\, h\; ))\, \equiv\;\, \exp\,\, (\,\, \frac {1} {4}
\:\, (\langle\, g\: ,\, g\; \rangle+\langle\; h\; ,\, h\; \rangle)\, )\: .
\quad\quad
\end {equation}
In fact, expectations over monomials of $\, a\, $ and $\, a^{\, *}\, $ can be expressed in terms of 
(\ref {propagatore Fock}) with the aid of Wick's theorem \cite {Wic50}, while those over 
monomials of $\, W\, $ are identified by (\ref {aspettazione esponenziale}) up to a 
phase factor, given by (\ref {commutatore W}), and the other expectations follow 
from (\ref {commutatore a W}).

The Gelfand-Naimark-Segal ($GNS$) construction \cite {Nai59} on a ${}^{*}$-algebra $\, \mathcal{A}\, $ 
and a linear (non-positive) functional $\, \omega\, $ proceeds as in the positive case, resulting in a 
non-degenerate indefinite vector space, a representation of $\, \mathcal{A}\, $ on it, with the 
$\, {}^{*}\, $ operation represented by the indefinite-space adjoint $\, {}^{\dagger}\, $, and 
expectations over a cyclic vector representing $\, \omega\, .\, $

The space $\, \mathscr {G}\, $ is obtained via the $\, GNS\, $ construction over $\, \omega_{\; F\; }^{} 
(\, \mathscr {A}_{\,\, ext}^{\,\, e.\, m.}\, )\, $ and its indefinite inner product is denoted by 
$\, \langle\, .\, ,.\, \rangle\, .\, $ 
In the sequel, $\, \mathscr {V}\, $ is regarded as a topological space with the weak 
topology $\, \tau_{\,\, w}^{}\, ,$ defined by the family of seminorms 
$\, p_{\, y}^{}\, (\, x\, )=\vert\langle\, y\, ,x\, \rangle\vert\, ,$ with 
$\, y\in\mathscr {V}.\, $
The domain $\, \mathscr {V}_{\;\, 0}^{}\, $ is given by 
\begin {equation}\label {definizione V_0} 
\mathscr {V}_{\;\, 0}^{}\, \equiv\,\, \mathscr {S\, }(\, \mathbb {R}^{\, 3}\, )
\, \otimes\; \mathscr {G}_{\;\, 0}^{}\, ,\nonumber 
\end {equation} 
with $\, \mathscr {G}_{\,\, 0}^{}\, $ obtained by performing the same construction described above for 
$\, \mathscr {G}\, ,\, $ for canonical and Weyl operators with test functions in $\, \mathscr {S}\, 
(\, \mathbb {R}^{\, 3}\, )\, .\, $ Since no confusion should arise, the indefinite inner product of 
$\, \mathscr {G}_{\,\, 0}^{}\, $ is again denoted by $\, \langle\, .\, ,.\, \rangle\, .\, $ 
$\, \mathscr {G}_{\;\, 0}^{}\, $ is weakly dense in $\, \mathscr {G}\, $ and $\, \mathscr {V}_{\;\, 0}^{}\, $ is 
weakly dense in $\, \mathscr {V}$ by density of the Schwartz space in $L^{\, 2\; }$ and Schwartz's inequality, 
applied to the explicit expression of the inner product.

The representation of the Weyl operators \emph {defines} the corresponding exponentials of the creation 
and annihilation operators on $\, \mathscr {G}\, ;\, $ such exponentials are also given by their series, 
which converge weakly on $\, \mathscr {G}\, $, a fact which is not needed in the following.
The above construction is suggested by the fact that the formal time-evolution operator defined by 
the Hamiltonian (\ref {hamiltoniana PFBR}) contains exponentials of the canonical variables of the 
soft-photon field.

We recall that indefinite-metric spaces obtained via a $\, GNS\, $ procedure, starting from a non-positive 
functional and a ${}^{*}$-algebra $\, \mathscr {A},\, $ are neither complete nor do they admit a unique 
completion.
In some generality, complete spaces can be introduced as Hilbert-space completions of vector spaces 
obtained through $\, GNS\, $ constructions.
While in general the Hilbert structure can be very relevant for the existence and control of limits (the role of 
Hilbert space structures in the formulation of models of indefinite metric quantum field theories has been 
discussed at length in \cite {MS80}), in the case of simple models, only involving polynomials and 
exponentials of fields, the use of a vector space looks simpler and even more intrinsic; a strong 
topology will only be needed for the formulation and control of uniqueness of 
time-evolution operators.

The standard positive scalar product on $\, \mathscr {G}\, ,\, $ obtained through the change of the sign of 
the summation over the space indices on the r.h.s. of (\ref {prodotto indefinito}), gives rise to the 
Hilbert space $\, \mathscr {H}\, $ introduced in \cite {HS09}. 

The free and full dynamics of the model are determined by isometries of $\, \mathscr {V}\, $ leaving 
$\, \mathscr {V}_{\,\, 0}^{}\, $ invariant and differentiable on $\, \mathscr {V}_{\,\, 0}^{}\, $ in the 
strong topology of $\, \mathscr {H},\, $ with derivative respectively given by  $\, H_{\; 0\, }^{}\, $ 
and $\, H_{\; \lambda }^{\,\, (PFBR )}\, $.\\
Notice that isometries $\, U\, $ of a non-degenerate indefinite space $\, \mathscr {Z}\, $ are determined by 
their restriction to a weakly dense subspace; in fact, by non degeneracy, $ U\, x\, ,$ $\, x\, \in\mathscr {Z}\, ,
\, $ is determined by 
$$
\langle\; y\, ,\, U\, x\, \rangle\, =\, \langle\; U^{\, -\, 1}\; y\, ,\,  x\, \rangle\, =\; \lim_{n\, }\,\, \langle\; 
U^{\, -\, 1}\; y\, ,\,  x_{\: n}^{}\, \rangle\, =\; \lim_{n\, }\,\, \langle\, y\, ,\, U\, x_{\: n}^{}\, \rangle\, ,  
$$
with $\, y \in \mathscr {Z},$ $x_{\, n}^{}$ in a weakly dense subspace. 
 
The uniqueness of the evolution operators is a consequence of the following observation, only 
requiring the existence of a positive scalar product $ (.\, ,.)$ majorizing the indefinite product,
$|\langle\, y\, , \, x\, \rangle| \leq (\, x\, ,\, x\, )^{\; 1\, /\, 2}\,\, (\, y\: ,\, y\, )^{\; 1\, /\: 2}\,\, $:
 
\begin {lemma}
If two one-parameter families of isometries $\, U\, (\, a\, )\, $ and $\, V\, (\, a\, )\, $ of a vector space 
$\, V_{\,\, 0}^{}\, ,$ endowed with a non-degenerate indefinite inner product $\, \langle\, .\, ,.\, 
\rangle\, $ which is majorized by a positive scalar product, are differentiable on $\, V_{\,\, 0}^{}$ 
in the corresponding strong Hilbert topology with the same derivative $\, -\, i\; H\, (\, a\, )\, ,\, $ 
then they coincide.
\end {lemma}
\begin {proof} 
Since strong differentiability implies strong continuity and hermiticity of $H\, (\, a\, )$ on $V_{\,\, 0}^{}\, ,\, $
one has, $\, \forall\; x\, ,\, y\, \in\, V_{\,\, 0}^{}\, $,  
\begin {eqnarray}
\frac {\; d} {d\; a\, }\;\, \langle\, x\, ,\; V\, (\, -\, a\, )\;\, U\, (\, a\, )\,\, y\, \rangle\, = 
\, \frac {\; d} {d\; a\, }\;\, \langle\; V\, (\, a\, )\,\, x\, , \, U\, (\, a\, )\,\, y\, \rangle
\nonumber\; \quad\quad\quad\quad\quad\quad\quad\\
=-\; i\;\, \langle\; V\, (\, a\, )\,\, x\, ,\,  H\, (\, a\, )\;\, U\, (\, a\, )\,\, y\, \rangle\,
+\, i\;\, \langle\, H\, (\, a\, )\;\, V\, (\, a\, )\,\, x\, ,\; U\, (\, a\, )\,\, y\, \rangle\, =\, 0\; .  
\nonumber
\end {eqnarray}
\end {proof} 

The dynamics of the model in the interaction representation is determined by an operator $\, U_{\; I}^{}:
\mathscr {V}\rightarrow\mathscr {V}$ such that the spectral component $U_{\,\, I\, ,\,\, \mathbf {v}}^{}\, ,$ 
obtained from the decomposition of $U_{\,\, I}^{}\, $ with respect to the joint spectrum of the components
of $\, \mathbf {v}\, ,\, $ is $\, \langle\, .\, ,.\, \rangle\; $- isometric for (almost) all $\, \mathbf {v}$ 
and leaves invariant a $\, \tau_{\; w}^{}\, $- dense subspace of $\, \mathscr {V}\, ,\, $ on which it 
is strongly differentiable with a time-derivative satisfying  
(\ref {equazione di evoluzione interazione}).
By employing formula (\ref {caso particolare formula BH}), one obtains a formal solution of the form
\begin {equation}\label {soluzione formale pauli fierz quadrivettoriale}
\quad\quad\quad U_{\,\, I\, ,\; \mathbf {v}}^{}\; (\, t\, )\, =\,\, \exp\,\, (\, -\; i\; e
\;\, (\; a^{\, \dagger\; }(f_{\; \tilde {v}\; }^{}(\, t\, ))\, +\, 
a\, (\, \overline {f\, }_{\tilde {v}\; }^{}(\, t\, ))\, )\, )\: ,\;\;
\end {equation}
\begin {equation}\label {funzione di coerenza PFBR}
\quad\quad\quad f_{\; \tilde {v}}^{\,\, \mu\; }(\, \mathbf {k}\, ,\; t\; )\, =\, 
\frac {\; \tilde {\rho\; }(\, \mathbf {k}\, )
\;\, {\tilde {v\; }}^{\mu}} {\sqrt {\; 2\;\, \omega_{\,\, \mathbf {k}}^{}}}\;\, 
\frac {\, e^{\; i\,\, \omega_{\: \mathbf {k}}^{}\; t\; }-\, 1\, } {i\,\,\, 
\omega_{\,\, \mathbf {k}}^{}}
\; \cdot\quad\;\;
\end {equation}
The operators (\ref {soluzione formale pauli fierz quadrivettoriale}) fulfill 
(\ref {equazione di evoluzione interazione}) in the strong topology, 
on $\, \mathscr {V}_{\;\, 0}^{}\, $, which is left invariant. 

Proceeding as for the $\, PFB\, $ model, we introduce the adiabatic mean and the renormalization 
counterterm,
\begin {eqnarray}\label {hamiltoniano PFBR rinormalizzato}
H_{\, \lambda\; ,\; R}^{\; (\, \epsilon\, )}\, =\,\, m\, +\frac {1} {2}\,\, m\,\, {\mathbf {v}}^{\; 2}\, +
\, H_{\; 0}^{\:\, e.\, m.\, }+\, e\;\,\, \tilde {v}\; \cdot\, A\,\, (\, \rho\, ,\, \mathbf {x}= 0\; )
\;\,\, e^{\, -\, \epsilon\,\, \vert\, t\, \vert\; }
\quad\;\nonumber\\
-\; \frac {3} {4}\,\,\, e^{\: 2}\;\, m\,\, z\;\, e^{\, -\, 2\:\, \epsilon\; \vert\, t\, \vert\; }
\equiv\,\, m\, +\frac {1} {2}\,\, m\,\, {\mathbf {v}}^{\; 2}\, +\, 
H_{\; 0}^{\:\, e.\, m.\, }+\, H_{\, int\, ,\: R}^{\; (\, \epsilon\, )}\,\, ,
\end {eqnarray}
and determine the corresponding evolution operator $\, U_{\,\, I\; ,\,\, \lambda\,\, }^{\; (\; \epsilon\, )}(\, t\, )\, $ 
in the interaction picture.
For positive times, $\, U_{\,\, I\: ,\,\, \lambda\,\, }^{\; (\, \epsilon\, )}(\, t\, )\, $ is of the form
\begin {equation}\label {operatore di evoluzione PFBR}
U_{\,\, I\; ,\,\, \lambda\,\, }^{\; (\, \epsilon\, )}(\, t\, )\, =\:\, \tilde {h}_{\; z}^{\; (\, \epsilon\, )}\, (\, t\, )
\,\,\, \exp\,\, (\, -\; i\; e\:\, (\: a^{\, \dagger\; }(\, f_{\,\, \tilde {v\, }}^{\,\, (\, \epsilon\, )}\, (\, t\, ))\, +\,
a^{}\: (\, \overline {f\, }_{\tilde {v\, }}^{\,\, (\, \epsilon\, )}\, (\, t\, ))\, )\, )\: ,
\end {equation}
with
\begin {equation}\label {funzioni di coerenza PFBR}
f_{\,\, \tilde {v\, },\; \mu}^{\,\, (\, \epsilon\, )\, }\, (\, \mathbf {k}\, ,\, t\, )\, = \, \frac {\; \tilde {\rho\,\, }(\, \mathbf {k}\, )
\,\,\, \tilde {v\, }_{\mu}^{}\, } {\sqrt {\; 2\;\, \omega_{\,\, \mathbf {k}}^{}}\, }\;\, 
\frac {\, e^{\; (\, i\,\, \omega_{\, \mathbf {k}}^{} -\; \epsilon\, )\;\, t}\, -\, 1\, } 
{i\;\, \omega_{\,\, \mathbf {k}}^{}-\, \epsilon}\, \cdot\;\;\;
\end {equation}
Considering that the above interaction Hamiltonian is Lorentz covariant only up to the second order in the 
four-velocity of the particle, it is not surprising that the above mass renormalization term, of the same form 
as that occurring in $\, QED\, ,\, $ only allows for the construction of the M\"{o}ller operators up to higher 
(fourth order) terms in the four-velocity.
We drop therefore the fourth order terms in $\, \tilde {h}_{\; z}\, $ and consider the (interaction representation) 
evolution defined by  (\ref{operatore di evoluzione PFBR}), (\ref{funzioni di coerenza PFBR}), with
\begin {equation}
\tilde {h}_{\; z}^{\,\, (\, \epsilon\, )}\; (\, t\, )\, \equiv\;\, \exp\,\, (\, -\; \frac {\, i\; e^{\, 2}} {2\; }
\;\, d^{\;\, (\, \epsilon\, )}\; (\, t\, )\, )\;\,\, \exp\,\, (\, -\; 3\,\, i\,\, e^{\: 2\;\, }m\,\, z\,\,\, 
\frac {\, e^{\, -\, 2\,\, \epsilon\:\, t}\, -\, 1\, } {8\,\, \epsilon}\,\, )\: .
\end {equation}
For $\, t<0\, ,\, $ one has to replace $\, \epsilon\, $ by $\, -\, \epsilon\, $ in the expressions above.\\
The asymptotic limits of such evolution operators are given by the following
\begin {theorem}\label {proposizione Moller Feynman PFBR}
The large-time limits and the adiabatic limit of the evolution operator (\ref {operatore di evoluzione PFBR}) 
exist on $\, \mathscr {V}_{\;\, 0}^{}\, $ and define M\"{o}ller operators as isometries of $\, \mathscr {V}\, $: 
\begin {eqnarray}\label {Moller Feynman PFBR}
\Omega_{\; \pm}^{\,\, (\, \lambda\, )\, }=\; \tau_{\; w}^{}-\lim_{\epsilon\; \rightarrow\; 0}\;
\lim_{t\,\, \rightarrow\, \mp\, \infty}\; U_{\,\, I\; ,\; \lambda\,\, }^{\; (\, \epsilon\, )}(\, -\; t\; )\,
\quad\quad\quad\quad\nonumber\\
=\; \tilde {h}_{\; \mp\, ,\; z\, }^{}\,\, \exp\,\, (\; i\; e\;\, [\,\, a^{\, \dagger\; }(\, f_{\,\, \tilde {v\, }\, }^{})+
\, a\; (\, \overline {f\, }_{\tilde {v\, }\, }^{})\, ]\, )\: ,\;\; 
\end {eqnarray}
\begin {equation}
\quad\quad\quad\quad\quad 
\tilde {h}_{\; \mp\, ,\; z\, }^{}=\, \lim_{\epsilon\; \rightarrow\; 0}\;\, \lim_{t\,\, \rightarrow\, \mp\, \infty}
\,\, \tilde {h}_{\; z}^{\; (\, \epsilon\, )}\; (\, t\, )\, =\; 1\, +\, O\,\, (\, \mathbf {v}^{\; 2}\; )\; ,
\quad
\end {equation}
\begin {equation}\label {f Moll PFBR}
f_{\; \tilde {v\, }}^{\; \mu\; }(\, \mathbf {k}\, )\, =
\, \frac {\; \tilde {\rho\,\, }(\, \mathbf {k}\, )\,\,\, \tilde {v\, }^{\, \mu}\, } 
{\sqrt {\; 2\;\, \omega_{\; \mathbf {k}}^{}}\, }\;\, \frac {i}
{\omega_{\; \mathbf {k}}^{}\, }\, \cdot\quad\quad\quad\quad\quad\quad
\end {equation}
\end {theorem}
\begin {proof}
By (\ref {commutatore W})-(\ref {aspettazione esponenziale}), the convergence of the coherence 
functions (\ref {funzioni di coerenza PFBR}) in $\, L^{\, 2}\, $ implies the weak convergence of the 
corresponding expectations on $\, \mathscr {G}_{\:\, 0}^{}\, $ of polynomials and exponentials of 
the smeared photon variables. 
$\, \Omega_{\; \pm}^{\,\, (\, \lambda\, )\, }$  are invertible and preserve the inner product 
$\, \langle\, .\, ,.\, \rangle\, $ as a consequence of the $\, GNS\, $ representation of 
(\ref {proprieta' W}),(\ref {commutatore W}). 
They define therefore isometries of $\, \mathscr {V}\, ,$ uniquely determined by their 
restrictions to $\, \mathscr {V}_{\,\, 0}^{}\, $.   
\end {proof}

As before, one can define a scattering operator 
\begin {equation}\label {matrice di scattering PFBR}
S_{\,\, \lambda}^{\,\, (PFBR\, )}\, =\;\, \tau_{\; w}^{}-\, \lim_{\epsilon\,\, \rightarrow\; 0}
\;\, \lim_{t\, ,\; t\, '\, \rightarrow\, +\, \infty}\;\, U_{\,\, I\; ,\,\, \lambda}^{\; (\, \epsilon\, )}\,\, (\, t\, )\;\,\, 
\mathscr {W}\;\,\, U_{\,\, I\; ,\,\, \lambda}^{\; (\, \epsilon\, )}\,\, (\; t\; '\; )\: 
\end {equation}
taking $\, \mathscr {W}\, $ as an isometric operator on $\, L^{\: 2}\; (\, \mathbb {R}^{\: 3}\, )\, \otimes\;
\mathscr {G}\, ,\, $ with the same role as in~(\ref {matrice di scattering PFB}); in the comparison with 
the diagrammatic expansion, its matrix elements are interpreted as the non-infrared 
contributions to the corresponding scattering process.

The expansion in powers of the electric charge of suitable matrix elements of the M\"{o}ller operators
of the $PFBR$ model reproduces qualitatively the infrared contributions of Dyson's power series, with 
modifications only due to the dipole approximation and to the non-relativistic limit.
For instance, consider the transition amplitude for the scattering between two single-particle states 
$\, \psi_{\; v}^{}\, $ and $\, \psi_{\; v\, '\; }^{},$ without external (massive) photons,
\begin {eqnarray}\label {elemento di matrice rel}
(\, S_{\; \lambda\,\, }^{\,\, (PFBR\, )}\, )_{\; v\, ' ,\,\, v\, }^{}\equiv\,\, \langle\; \Omega_{\; -}^{\: (\, \lambda\, )\; }
(\, \psi_{\; v\, '}^{}\otimes\, \Psi_{\, F}^{}\, )\; ,\; \mathscr {W}\,\,\, \Omega_{\; +}^{\, (\, \lambda\, )}\,\, 
(\, \psi_{\; v}^{}\, \otimes\, \Psi_{\, F}^{}\, )\, \rangle
\nonumber\\
\quad\quad\quad\quad\quad\quad\quad\quad\quad\,\,\,\, =
\langle\; \Omega_{\; -\, ,\:\, \mathbf  {v}\, '}^{\: (\, \lambda\, )\; }\; \Psi_{\, F}^{}\,\, ,
\,\, \Omega_{\; +\, ,\:\, \mathbf  {v}}^{\; (\, \lambda\, )}\,\, \Psi_{\, F}^{}\; \rangle\;\,\,
\mathscr {W}_{\,\, v\, ',\:\, v}^{}\; ,
\end {eqnarray}
where $\, \Omega_{\; +\, ,\,\, \mathbf {v}}^{\: (\, \lambda\, )}$ is obtained from the decomposition of 
$\, \Omega_{\; +}^{\, (\, \lambda\, )}$ with respect to the joint spectrum of the components of 
$\, \mathbf  {v}\, .\, $
The radiative soft-photon corrections to the basic process are reproduced by the term in brackets 
in the second line of  (\ref {elemento di matrice rel}).
The exponentiation of the low-energy radiative corrections is thus a consequence of (\ref {caso 
particolare formula BH}) and the compatibility of the dipole approximation with the 
renormalization procedure, with the limitation to the second order in the particle 
four-velocity discussed above, is also displayed.

However, the transition amplitude (\ref {elemento di matrice rel}) \emph {is not} equal to the corresponding 
expression in the Coulomb gauge, given by $(\, S_{\, \lambda\,\, }^{\,\, (PFB\, )}\, )_{\; v\, ',\; v}^{}\, ,$ namely, it is 
not gauge invariant. 
For clarity of exposition, we point out that statements about the property of gauge-independence are not 
referred in this work to the concept of gauge invariance in its broad sense.
Rather, since we employ the interaction representation and the adiabatic approximation, the asymptotic states 
obey a free evolution; the notion of gauge-independence of the transition amplitudes between physical 
scattering states in our models thus coincides with that of a free theory, namely with ``the cancellation 
of the contributions from unphysical polarizations'' to such amplitudes.

A first indication that problems may arise comes from the fact that the dipole approximation prevents 
the local conservation of the electric charge in a gauge involving four independent photon degrees of 
freedom; as a consequence, the standard argument explaining the cancellation of the contributions 
from longitudinal and scalar photons to the Feynman amplitudes no longer applies, since it requires 
the free-field character of $\, \partial\, \cdot A\, $, which in turn relies on the continuity equation.
The occurrence of infrared effects, due to the lack of (local) charge-conservation, can be explicitly 
checked; a simple calculation yields in fact
\begin {eqnarray}\label {ampiezze approssimate}
\langle\; \Omega_{\; -\, ,\:\, \mathbf  {v}\, '}^{\, (\, \lambda\, )\; }\; \Psi_{\, F}^{}\; ,\;
\Omega_{\; +\, ,\:\, \mathbf  {v}}^{\; (\, \lambda\, )}\,\, \Psi_{\, F}^{}\, \rangle\, =\;
\exp\; (\, -\, \frac {\, e^{\: 2}} {4\; }\;\, \vert\; \mathbf {v}\; '-\, \mathbf {v}
\; \vert^{\,\, 2}\, \int\; \frac {\, d^{\,\, 3\, }k} {\omega_{\; \mathbf {k}}^{\,\, 3}}\,\,
\, {\tilde {\rho\; }}^{2}\, (\, \mathbf {k}\, )\, )
\; \nonumber\\
\neq\: \exp\; (\, -\, \frac {\, e^{\: 2}} {6\, }\;\, \vert\; \mathbf {v}\; '-\, \mathbf {v}\; \vert^{\,\, 2}\, \int\; 
\frac {\, d^{\,\, 3\, }k} {\omega_{\; \mathbf {k}}^{\,\, 3}}\,\,\, {\tilde {\rho\; }}^{2}\, (\, \mathbf {k}\, )\, )
\; .
\end {eqnarray}
An extra factor $\, 3/2\, $ thus arises in the first exponent on the r.h.s. of (\ref {ampiezze approssimate}) 
with respect to the exponent on the second line, which is obtained by evaluating the scalar product 
$(\; \Omega_{\; -\, ,\,\, tr\, ,\; \mathbf  {v}\, '}^{\; (\, \lambda\,)\, }\, \Psi_{\, F}^{}\, ,\,
\Omega_{\; +\, ,\,\, tr\, ,\,\, \mathbf  {v}}^{\; (\, \lambda\, )}\; \Psi_{\, F}^{}\, )$ 
involving the M\"{o}ller operators of the $\, PFB\, $ model; 
consequently one has a different behaviour for the vanishing of (\ref {elemento di matrice rel}) in the 
limit $\, \lambda\rightarrow 0\, ,$ with respect to the corresponding Coulomb-gauge amplitude. 
The extra factor arises basically from the fact that the space components of the coherence 
functions (\ref{f Moll PFBR}) are of first order in the velocity, while only two components appear in 
(\ref {limiti forti}).

The relevant point is that the lack of gauge-invariance of (\ref {elemento di matrice rel}) depends neither 
on the dropping of the second order term in the interaction Hamiltonian nor on the approximation 
adopted in the computation of the evolution operator (\ref {operatore di evoluzione PFBR}), but is 
rather a difficulty common to all four-vector gauge quantizations in the presence of a dipole
approximation. 

In order to trace back the origin of such effects, it suffices to notice that the expansion of Dyson's 
$\, S\: $- matrix elements for, say, the scattering of an electron by a potential contains, in the 
non-relativistic limit, powers of the scalar photon term 
$(\: p\; '^{\,\, 0}/\, (\: p\; '\cdot\; k\, )-p\, ^{\, 0}/\, (\: p\, \cdot\, k\, )\, )^{\,\, 2}\simeq\omega_{\; \mathbf {k}}^{\; -\, 2\;\, }
[\, (\mathbf {\: v}\, '-\mathbf {\, v\; })\, \cdot\, \hat {\mathbf {k}}\,\, ]^{\,\, 2\; }, $ which do not appear in the 
presence of the dipole approximation; the residual contributions to the corresponding scattering amplitudes 
are thus due to the longitudinal photons. 

In \cite {HS09}, it has been pointed out that the dipole approximation prevents a straightforward 
application of the Gupta-Bleuler procedure, since the $\, \partial\, \cdot A\, $ field no longer obeys a 
free evolution.
As stressed in the above discussion, the difficulties with such an approximation are indeed not limited to 
the implementation of the Gupta-Bleuler condition, but are even more basic; spurious gauge-dependent 
contributions, also relevant in the infrared regime, are in fact introduced into the M\"{o}ller operators, 
thus not allowing to reproduce substantial features of the perturbative treatment of soft-photon effects in a 
local and covariant gauge. 
The discrepancy with the perturbative results extends \cite {Zer09} to the explicit expressions of the 
infrared-finite inclusive cross-sections~\cite {Wei95, JR76}.

As it will be shown in the forthcoming Section, such difficulties do not arise in models based on a 
Bloch-Nordsieck approximation.

\section {Bloch-Nordsieck Models}
\label {sect:2}

In the present Section we introduce hamiltonian models based on an approximation first devised by Bloch and 
Nordsieck \cite {BN37}, which turns out to amount to a first-order expansion around a fixed four-momentum 
of each charged particle, with respect to the energy-momentum transfer. 
Since this fact is not transparent in the original treatment, a brief discussion may be useful.\\
Consider the one-particle Dirac Hamiltonian with minimal coupling,
\begin {equation}
\quad\quad\;  
H\, =\,\, \mathbf {\alpha}\; \cdot\, (\,\, \mathbf {p}\, -\, e\,\, \mathbf {A}\; )\, +\, \beta\,\, m\; +
\, e\;\, A^{\,\, 0\; }\equiv\; H_{\, D\, }^{}-\, e\;\,\, \mathbf {\alpha}\, \cdot\, \mathbf {A}\; +\, 
e\,\, A^{\; 0}\; ,\, 
\end {equation}
and an eigenstate $\psi_{\: + ,\; p\; }^{}(\, x\, )=e^{\, -\, i\,\, p\, \cdot\, x\,\,\, }u_{\; r}^{}\, (\, \mathbf {p}\, )\, $ 
of $\, H_{\, D}^{}$ with momentum $\mathbf {p}\, $ and positive energy $\, E_{\: \mathbf {p}}^{}\, $, 
$u_{\; r}^{}\, (\, \mathbf {p}\, )\, $ being a spinor with helicity $\, r\, .\, $\\ 
Let $\, u_{\; r}^{}\, (\, \mathbf {p}\, )= u_{\,\, r}^{}\, (\, \mathbf {p}_{\; 0}^{}\, )\, +\, O\: (\, \mathbf {p}-
\mathbf {p}_{\; 0}^{}\, )\, ;\, $ by the algebraic relations for Dirac's matrices one finds
\begin {eqnarray}\label {hamiltoniano BN dall'espansione}
H_{\, D\;\,\, }^{}\psi_{\; 0\, ,\; p\,\, }^{}(\, x\, )\, =\; [\,\, \mathbf {v}\, \cdot\, \mathbf {p}\; +
\,\sqrt {\, 1-\, \mathbf {v}^{\: 2\, }}\;\,\, m\,\, ]\;\,\, \psi_{\; 0\, ,\; p\,\, }^{}(\, x\, )\, 
+\, O\; (\, \mathbf {p}-\mathbf {p}_{\; 0}^{}\, )\; ,
\\\quad\quad\quad\quad 
\; \psi_{\; 0\, ,\,\, p\; }^{}(\, x\, )\, \equiv\; e^{\, -\, i\,\, p\, \cdot\, x}\,
\,\, u_{\; r}^{}\, (\, \mathbf {p}_{\; 0}^{}\, )\; ,\; \mathbf {v}\, 
\equiv\, E_{\; \mathbf {p}_{\, 0}^{}\, }/\, \mathbf {p}_{\; 0}^{}\, . 
\nonumber
\end {eqnarray}
The $\, \mathbf {v}\; $- dependent terms on the r.h.s of (\ref {hamiltoniano BN dall'espansione}) could also 
be obtained by formally replacing the matrices $\, \mathbf {\alpha}\, $ and $\, \beta\, $ in $\, H_{\, D}^{}\, $ 
respectively by the (diagonal in the spinor indices) matrices $\, \mathbf {v}\, $ and 
$\, \sqrt {\, 1-\mathbf {v}^{\; 2}}\, .\, $
Although this result may seem to rely on the linearity of $\, H_{\, D}^{}\, $ with respect to the $\, \alpha\, $ 
matrices, it is indeed more general; for instance, it would also be obtained by performing a similar 
expansion on the eigenvalue equation for the Klein-Gordon Hamiltonian.

According to the above discussion, we introduce the models defined by the Hamiltonians, respectively in 
the Coulomb gauge and in the $\, FGB\, $ gauge,
\begin {equation}\label {hamiltoniano coulomb bn}
H_{\; \lambda}^{\,\, C}\, = \; {\mathbf {p}}\, \cdot\, \mathbf {v}\; +
\, H_{\; 0\; ,\,\, tr\; }^{\; e.\, m.\, } - 
\, e\;\; \mathbf {v}\, \cdot\mathbf {A}_{\, tr\; }^{}(\, \rho\, ,\,  {\mathbf {x}}\, )\, \equiv\; 
H_{\; 0\;\, }^{\; C}+\, H_{\, int\; }^{\; C}\, ,
\end {equation}
\begin {equation}\label {hamiltoniano Feynman}
H_{\; \lambda}^{\; F}\, =\; {\mathbf {p}}\, \cdot\, \mathbf {v}\, +\, H_{\; 0\,\, }^{\,\, e.\, m.}+
\, e\;\,\, v\, \cdot A\,\, (\, \rho\, ,\: {\mathbf {x}}\, )\, \equiv\; H_{\,\, 0\;\, }^{\; F}+\, H_{\, int\,\, }^{\; F}\, ,\quad
\end {equation}
with $\, v\equiv(\, 1\, ,\, \mathbf {v}\, )\, ,\, \mathbf {v}\, $ a triple of self-adjoint operators in 
a Hilbert space, to be identified as the observable corresponding to the asymptotic velocity of the 
particle. 
They commute with the Weyl algebra $\, \mathscr {A}_{\; ch}^{}\, $ generated by the canonical variables of 
the electron and with the polynomial algebras generated in the Feynman gauge by the photon canonical 
variables.
Due to the appearance of the operators $\, \mathbf {x}\, ,\, $ the above Hamiltonians \emph {do commute 
with space translations}. 
The e.m. potentials occurring in (\ref {hamiltoniano coulomb bn}), (\ref {hamiltoniano Feynman}) will be 
interpreted as describing soft-photon degrees of freedom. 
 
First we consider the model formulated in the Coulomb gauge.
With the same notation as in Section \ref {sect:1}, the Hilbert space is 
$$
\mathscr {H}\, \equiv\,\, L^{\; 2\; }(\, \mathbb {R}^{\, 3}\, )\, \otimes\, L^{\; 2\; }(\: \mathbb {R}^{\, 3}\; 
d^{\,\, 3\, }v\, )\, \otimes\, \mathscr {F\, }\equiv\; \int\, d^{\,\, 3\, }v\;\, {\mathscr {H}}_{\,\, \mathbf {v}}^{}\, , 
$$
$\, \mathbf {v}\, $ is a multiple of the identity on~$\, {\mathscr {H}}_{\,\, \mathbf {v}}^{}\, $ and the model 
can be studied at fixed~$\, \mathbf {v}\, ,\, $ with Hamiltonian $\, H_{\; \lambda}^{\,\, (\, \mathbf{v}\, )}\, $ given 
by (\ref{hamiltoniano coulomb bn}).

Proceeding in a similar way as in the proof of the self-adjointness of the Pauli-Fierz Hamiltonian, 
one finds that $\, H_{\; \lambda}^{\;\, C\, ,\,\, (\, \mathbf{v}\,) }\, $ is e.s.a. on $\, D_{\; 0}^{}\subset 
{\mathscr {H}}_{\,\, \mathbf {v}}^{}\, ;\, $ existence and uniqueness of the dynamics thus follow 
by Stone's theorem. 
The equations of the motion in the interaction representation, with the aid of  
(\ref {caso particolare formula BH}), give 
\begin {equation}
\mathscr {U}_{\:\, I\, ,\,\, tr\, ,\; \lambda}^{}\; (\, t\, )\, =\; c_{\,\, 1\; }^{}(\, t\, )\,\,\, \exp\,\, (\; i\; e
\,\, (\; a_{\; tr}^{\; *}\, (\, f_{\,\, \mathbf {v}\, \mathbf {x}\; }^{}(\, t\, ))\, +\, a_{\; tr}^{}
\, (\: \overline {f\, }_{\mathbf {v}\, \mathbf {x}\; }^{}(\, t\, ))\, )\, )\; ,
\end {equation}
\begin {eqnarray}
c_{\,\, 1}^{}\; (\, t\, )\, =\;\, \exp\,\, (\;\, \frac {\; i\,\, e^{\, 2}\:\, 
{\mathbf {v}}^{\; 2}} {3\; }\;\; d_{\,\, 1}^{}\,\, (\, t\, )\, )\: , 
\quad\quad\quad\quad\quad\quad\quad\; \nonumber\\
d_{\; 1}^{}\; (\, t\, )\, =\, \int\,\, \frac {\, d^{\,\, 3\, }k} {\omega_{\,\, \mathbf {k}}^{}
\,\,\, v\, \cdot\, k\, }\;\,\, {\tilde {\rho\,\, }}^{2\,\, }(\, \mathbf {k}\, )\;\, (\,\, t\, -\frac {\, \sin\; v\, \cdot\, k\;\, t} 
{\;\;\, v\, \cdot\, k}\;\, )\: ,
\nonumber
\end {eqnarray}
\begin {equation}
\quad\quad\quad f_{\,\, \mathbf {v}\, s\; \mathbf {x}}^{}\; (\, \mathbf {k}\, ,\; t\, )\, =
\,\, e^{\, -\, i\;\, \mathbf {k}\; \cdot\; \mathbf {x}}\;\,\, \frac {\; \tilde {\rho\,\, }
(\, \mathbf {k}\, )} {\sqrt {\; 2\,\,\, \omega_{\,\, \mathbf {k}}^{}}\, }\;\,\, 
\mathbf {v}\, \cdot\, \mathbf{\epsilon}_{\; s}^{}\; (\, \mathbf {k}\, )
\,\,\, \frac {\, e^{\,\, i\;\, v\; \cdot\; k\;\; t}\, -\, 1}
{i\;\, v\, \cdot\, k\, }\, \cdot
\end {equation}
With the same motivations and following the same treatment as in the first Section, we introduce the 
adiabatic renormalized Hamiltonian 
\begin {eqnarray}\label {hamiltoniano fp con controtermine}
\quad\quad H_{\; \lambda\; ,\; R}^{\; (\, \mathbf {v}\, )\; ,\,\, (\, \epsilon\, )}\; =\,\, H_{\,\, 0\;\, }^{\; (\, \mathbf {v}\, )}\, +
\, H_{\; int\; }^{\,\, (\, \mathbf {v}\, )\,\, }\;\, e^{\, -\, \epsilon\,\, \vert\, t\, \vert\,\, }+
\; e^{\; 2}\,\,\, z_{\,\, 1}^{}\, (\, v\, )\;\,\, {\mathbf {v}}^{\; 2}\;\; 
e^{\, -\, 2\;\, \epsilon\,\, \vert\, t\, \vert}
\nonumber\\ 
\equiv\; H_{\,\, 0\;\, }^{\; (\, \mathbf {v}\, )}\, +\;
H_{\; int\, ,\; R}^{\; (\, \mathbf {v}\, )\, ,\,\, (\, \epsilon\, )\;\, },\,
\end {eqnarray}
\begin {equation}\label {controtermine BN Coulomb}
z_{\,\, 1}^{}\, (\, v\, )\, =\; \frac {1} {3}\,\,\, \int\; 
\frac {\, d^{\,\, 3\, }k} {\omega_{\; \mathbf {k}}^{}}
\;\, \frac {\, \tilde {\rho\; }^{2\; }(\, \mathbf {k}\, )} 
{v\, \cdot\, k\, }\, \cdot
\end {equation}
Kato's result on time-dependent Hamiltonians applies as in Section \ref {sect:1} and, using 
(\ref {caso particolare formula BH}), one obtains the corresponding evolution operator in 
the interaction representation; for positive times
\begin {equation}
\mathscr {U}_{\,\, I\, ,\,\, tr\, ,\; \lambda\,\, }^{\; (\, \epsilon\, )}(\, t\, )\, =\,\, h_{\,\, z_{\, 1}^{}}^{\,\, (\, \epsilon\, )}\, (\, t\, )
\,\,\, \exp\,\, (\,\, i\,\, e\;\, (\; a_{\; tr}^{\; *}\, (\, f_{\,\, \mathbf {v}\, \mathbf {x}}^{\; (\, \epsilon\, )}\, (\, t\, ))\, +
\; a_{\,\, tr}^{}\, (\, \overline {f\, }_{\mathbf {v}\, \mathbf {x}}^{\; (\, \epsilon\, )}\, (\, t\, ))\, )\, )\; ,
\end {equation}
\begin {equation}\label {esponenziale regolarizzato}
h_{\; z_{\, 1}^{}}^{\; (\, \epsilon\, )}\, (\, t\, )\, =\,\, \exp\; (\; \frac {\, i\,\, e^{\, 2}
\; {\mathbf {v}}^{\; 2}} {3\, }\,\,\, d_{\; 1}^{\,\, (\, \epsilon\, )}\, (\, t\, )\, )\; 
\, \exp\; (\; i\,\, e^{\, 2}\,\, z_{\; 1}^{}\, (\, v\, )\,\,\, {\mathbf {v}}^{\; 2}
\;\, \frac {\, e^{\, -\, 2\,\, \epsilon\,\, t}\, -\, 1\, } {2\,\, \epsilon\, }\,\, )\; ,
\end {equation}
\begin {eqnarray}
d_{\; 1}^{\,\, (\, \epsilon\, )}\, (\, t\, )\, =\, -\, \int\,\, \frac {\, d^{\,\, 3\, }k} 
{\omega_{\,\, \mathbf {k}}^{}\; (\, (\, v\, \cdot\, k\, )^{\,\, 2}\, +\, 
\epsilon^{\; 2}\, )}\;\,\, {\tilde {\rho\; }}^{2\; } 
(\, \mathbf {k}\, )\;\, (\,\, e^{\, -\, \epsilon\,\, t\, }
\;\, \sin\; v\, \cdot\, k\,\,\, t\,\,\, 
\nonumber\\
\quad\quad\quad\;  
+\; \frac {\, v\, \cdot\, k} {2\,\, \epsilon}\;\,\, 
(\,\, e^{\, -\, 2\,\, \epsilon\,\, t\, }\, -\, 1\; )\, )\; ,
\end {eqnarray}
\begin {equation}
f_{\,\, \mathbf {v}\; s\,\, \mathbf {x}}^{\; (\, \epsilon\, )}\; (\, \mathbf {k}\, ,\; t\, )\, =
\,\, e^{\, -\, i\;\, \mathbf {k}\, \cdot\; \mathbf {x}}\,\,\,\, \frac {\; \tilde {\rho\,\, }
(\, \mathbf {k}\, )} {\sqrt {\; 2\;\, \omega_{\,\, \mathbf {k}}^{}}}\;\, 
\mathbf {v}\, \cdot\, \mathbf{\epsilon}_{\; s}^{}\, (\, \mathbf {k}\, )
\,\,\, \frac {\, e^{\,\, i\,\, v\; \cdot\; k\;\, t\,\, -\,\, \epsilon\,\, t}\, -\, 1}
{i\;\, v\, \cdot\, k\, -\, \epsilon}\, \cdot
\end {equation}
The results for $\, t<0\, $ are obtained by replacing $\, \epsilon\, $ with $\, -\, \epsilon\, $ in the expressions 
above.
One can then prove the following 
\begin {theorem}
The evolution operator in the interaction representation corresponding to the Hamiltonian 
(\ref {hamiltoniano fp con controtermine}), with the counterterm given by  
(\ref {controtermine BN Coulomb}), admits asymptotic limits, yielding the 
M\"{o}ller operators in the Coulomb gauge
\begin {eqnarray}\label {Moller BN Coulomb}
\Omega_{\; \pm\, ,\,\, tr}^{\; (\, \lambda\, )}\, =\;\, s\; -\; \lim_{\epsilon\; \rightarrow\; 0}\;\, \lim_{t\,\, \rightarrow\, \mp\; \infty}
\, \mathscr {U}_{\,\, I\, ,\,\, tr\, ,\; \lambda\,\, }^{\; (\, \epsilon\, )}(\, -\; t\; )
\quad\quad\quad\quad\quad\quad\quad\quad\nonumber\\
=\;\, \exp\,\, (\, -\; i\; e\,\, \sum_{s\; }\;\,\, [\,\, a_{\,\, s}^{\,\, *\; }
(\, f_{\,\, \mathbf {v}_{\, \mp}^{}\, s\,\, \mathbf {x}}^{}\, )\, +\, 
a_{\; s\,\, }^{}(\, \overline {f\, }_{\mathbf {v}_{\, \mp}^{}\, s\,\, 
\mathbf {x}}^{}\, )\, ]\, )\; ,\;\;
\end {eqnarray}
\begin {equation}\label {coerenza Coulomb}
f_{\,\, \mathbf {v}\; s\,\, \mathbf {x}}^{}\; (\, \mathbf {k}\, )\, =\;\, e^{\, -\, i\;\, \mathbf {k}\, \cdot\; \mathbf {x}}
\,\,\,\, \frac {\; \tilde {\rho\,\, }(\, \mathbf {k}\, )} {\sqrt {\; 2\;\, \omega_{\,\, \mathbf {k}}^{}}}
\,\,\, \frac {\; i\,\,\, \mathbf {v}\, \cdot\, \mathbf {\epsilon}_{\; s}^{}\, (\, \mathbf {k}\, )}
{v\, \cdot\, k\, }\, \cdot
\end {equation}
The $\, t\, \rightarrow\pm\, \infty\, $ and $\, \epsilon\, \rightarrow\, 0\, $ limits in (\ref {Moller BN Coulomb}) 
exists in the strong topology. 
\end {theorem}
\begin {proof}
The control of the limits is very similar to the case discussed in Proposition \ref {proposizione Moller PFB}.
\end {proof}
\ni
Scattering operators can be introduced as before in terms of operators acting on the ``asymptotic velocity 
Hilbert space'' $\, L^{\; 2\,\, }(\, \mathbb {R}^{\, 3}\,\, d^{\; 3\, }v\, )\, $.

We shall now turn to the analysis of the model defined by the Hamiltonian (\ref {hamiltoniano Feynman}), 
which will be also referred to as four-vector $\, BN\, $ model. 
Concerning the problems posed by the absence of a positive scalar product, we adopt the same choices 
as in Section \ref {sect:1}, following the same steps with identical results.
The construction of the space of the model demands some modifications due to the effects of translation 
invariance, which requires the introduction of e.m. canonical and Weyl operators, fulfilling equations of 
the same form as (\ref {proprieta' W})-(\ref {commutatore a W}), smeared with test functions of the 
form
\begin {equation}\label {funzioni test}
g_{\; \mathbf {x}}^{}\, (\, \mathbf {k}\, )\, \equiv\;\, \cos\; (\, \mathbf {k}\cdot
\mathbf {x}\, )\;\,\, f\, (\, \mathbf {k}\, )\: ,\,\, h_{\: \mathbf {x}}^{}\, (\, \mathbf {k}\, )\, \equiv\, -\, 
\sin\; (\, \mathbf {k}\cdot\mathbf {x}\, )\;\,\, f\, (\, \mathbf {k}\, )\; ,
 \end {equation}
with $\, f^{\: \mu}\, $ real valued and square integrable. 

The $\, GNS\, $ construction over such an algebra and the Fock vacuum yields an indefinite space 
$\, \mathscr {G}^{\; BN}\, $ and a weakly dense domain $\, \mathscr {G}_{\,\, 0}^{\: BN}\, ;\, $ 
the tensor product with the same particle space as for the Coulomb-gauge version of the 
model gives the indefinite-metric space 
$$
\mathscr {V}^{\; BN\, }\equiv\,\, L^{\; 2\,\, }(\, \mathbb {R}^{\, 3}\, )\, \otimes\, L^{\, 2}\; 
(\, \mathbb {R}^{\, 3}\; d^{\; 3\, }v\, )\, \otimes\, \mathscr {G}^{\; BN}
$$
and the weakly dense subspace
$$\mathscr {V}_{\;\, 0}^{\; BN\, }\equiv\,\, \mathscr {S\; }(\, \mathbb {R}^{\, 3}\, )\, \otimes\, 
\mathscr {S\; }(\, \mathbb {R}^{\, 3}\, ) \, \otimes \, \mathscr {G}_{\;\, 0}^{\; BN}\, ;
\quad\quad$$ 
$ \mathscr {V}^{\, BN}\, $ and $\, \mathscr {V}_{\;\, 0}^{\; BN}\, $ will be decomposed 
on the spectrum of $\, \mathbf{v}\, $ as above.
 
Isometric evolution operators on $\, \mathscr {V}_{\;\, 0}^{\; BN}\, $ are constructed as in Section 
\ref {sect:1} and are unique in the same sense. 
As for the Coulomb gauge, the model can be studied at fixed $\, {\mathbf {v}}\, .\, $  
The evolution operator in the interaction representation is given by 
\begin {equation}\label {operatore formale}
\mathscr {U}_{\,\, I\,\, }^{}(\, t\, )\, =\;\, c_{\,\, 2\,\, }^{}(\, t\, )\,\,\, \exp\,\, (\, -\; i\; e
\;\, (\; a^{\, \dagger}\, (\, f_{\; v\: \mathbf {x}}^{}\, (\, t\, ))\, +\,
a\; (\; \overline {f\, }_{v\: \mathbf {x}}^{}\, (\, t\, ))\, )\, )\; ,
\end {equation}
\begin {eqnarray}
c_{\,\, 2\,\, }^{}(\, t\, )\, =\;\, \exp\,\, (\, -\; \frac {\, i\,\, e^{\, 2}\,\, v^{\; 2}} {2\; }
\;\,\, d_{\,\, 1\,\, }^{}(\, t\, )\, )\; ,
\quad\quad\quad\quad\quad \nonumber\\
f_{\; v\; \mathbf {x}}^{\; \mu}\; (\, \mathbf {k}\, ,\: t\, )\, =\;\, e^{\, -\, i\;\, \mathbf {k}\, \cdot\; \mathbf {x}}\;\,\, 
\frac {\; \tilde {\rho\; }(\, \mathbf {k}\, )\;\, v^{\; \mu}\, } {\sqrt {\; 2\,\,\, \omega_{\,\, \mathbf {k}}^{}}\; }
\;\, \frac {\, e^{\; i\;\, v\,\, \cdot\,\, k\,\,\, t}\, -\, 1\, } {i\,\,\, v\, \cdot\, k}\, \cdot\;\; \nonumber
\end {eqnarray}
In order to construct M\"{o}ller operators, we introduce the renormalized Hamiltonian and the adiabatic factor,
\begin {eqnarray}\label {hamiltoniano Feynman rinormalizzato}
H_{\; \lambda\; ,\; R}^{\,\, (\, v\, )\,\, ,\,\, (\, \epsilon\, )\; }=\;\, \mathbf {p}\, \cdot\, \mathbf {v}\,\, +
\, H_{\; 0\; ,\,\, \lambda}^{\;\, e.\, m.}\, +\, e\;\,\, v\; \cdot\, A\,\, (\, \rho\; ,\, \mathbf {x}\; )\;
\,\, e^{\, -\, \epsilon\,\, \vert\, t\, \vert}\, 
\nonumber\\
-\; e^{\; 2}\,\,\, z_{\; 2\,\, }^{}(\, v\, )\;\,\, v^{\; 2}\,\,\, e^{\, -\, 2\;\, \epsilon\,\, \vert\, t\, \vert}\,\,
=\; H_{\,\, 0\; ,\,\, \lambda}^{\,\, (\, v\, )}\, +\, H_{\; int\, ,\; R}^{\,\, (\, v\, )\; ,\,\, (\, \epsilon\, )}\; ,
\end {eqnarray}
\begin {equation}\label {coefficiente controtermine covariante BN}
z_{\; 2\,\, }^{}(\, v\, )\, =\,\, \frac {1} {2}\;\, \int\,\, \frac {\, d^{\,\, 3\, }k}
{\omega_{\,\, \mathbf {k}}^{}}\;\, \frac {\; \tilde {\rho\,\, }^{2\,\, }
(\, \mathbf {k}\, )} {v\, \cdot\, k\, }\, =\; \frac {3} {2}\;\,\, 
z_{\,\, 1}^{}\; (\, v\, )\: .
\end {equation}
For $\, t>0\, $, the corresponding evolution operator in the interaction representation is
\begin {equation}\label {operatore di evoluzione BN rinormalizzato}
\mathscr {U}_{\,\, I\; ,\,\, \lambda}^{\,\, (\, \epsilon\, )}\; (\, t\, )\, =
\,\, h_{\; z_{\: 2}^{}}^{\,\, (\, \epsilon\, )}\, (\, t\, )\,\,\, \exp\,\, (\,\, i\; e\,\, 
(\, a^{\, \dagger}\, (\, f_{\; v\; \mathbf {x}}^{\,\, (\, \epsilon\, )}\, (\, t\, ))\, +
\, a^{}\, (\, \overline {f\, }_{v\; \mathbf {x}}^{\; (\, \epsilon\, )\; }
(\, t\, ))\, )\, )\; ,
\end {equation}
\begin {eqnarray}
h_{\; z_{\, 2}^{}}^{\; (\, \epsilon\, )}\; (\, t\, )\, \equiv\;\, \exp\,\, (\, -\; \frac {\, i\,\, e^{\: 2}\,\, v^{\; 2}} 
{2\; }\;\,\, d_{\,\, 1\;\; }^{\;\, (\, \epsilon\, )}\, (\, t\, )\, )\,\,
\quad\quad\quad\quad\quad\quad\quad\quad\nonumber\\
\times\; \exp\; (\, -\, i\,\, e^{\: 2}\;\, z_{\,\, 2}^{}\; (\, v\, )\;\, v^{\; 2}\;\, 
\frac {\, e^{\, -\, 2\,\, \epsilon\,\, t}\, -\, 1\, } {2\,\, \epsilon\, }\; )\; ,\quad\quad
\end {eqnarray}
\begin {equation}
f_{\; v\; \mathbf {x}}^{\,\, (\, \epsilon\, )\, ,\,\, \mu}\; (\, \mathbf {k}\, ,\; t\, )\, =
\;\, e^{\, -\, i\;\, \mathbf {k}\; \cdot\; \mathbf {x}}\;\,\, \frac {\; \tilde {\rho\,\, }
(\, \mathbf {k}\, )\,\,\, v^{\; \mu}\, } {\sqrt {\; 2\,\,\, 
\omega_{\,\, \mathbf {k}}^{}}\; }\;\, 
\frac {\, e^{\; i\,\, v\; \cdot\; k\;\, t\;\, -\; \epsilon\;\, t}\; -\, 1\, } 
{i\,\,\, v\, \cdot\, k\, -\, \epsilon}
\, \cdot
\end {equation}
The results holding for $\, t<0\, $ are obtained by replacing $\, \epsilon\, $ by $\, -\, \epsilon\, $ in the 
expressions above.
The asymptotic limits are controlled by the following
\begin {theorem}
The large-time limits and the adiabatic limit of the evolution operator 
(\ref {operatore di evoluzione BN rinormalizzato}), defining the 
M\"{o}ller operators of the four-vector $\, BN\, $ model, exist in 
the weak topology and define isometries in
$\, \mathscr {V}^{\; BN} $, given by
\begin {eqnarray}\label {Moller Feynman BN}
\Omega_{\; \pm}^{\: (\, \lambda\, )\, }\equiv\;\, \tau_{\; w}^{}\, -\lim_{\epsilon\; \rightarrow\; 0}
\;\, \lim_{t\,\, \rightarrow\,\, \mp\, \infty}\;\, \mathscr {U}_{\,\, I\; ,\,\, \lambda}^{\,\, (\, \epsilon\, )}\,\, (\, -\; t\, )\, \equiv
\,\, \tau_{\; w}^{}\, -\lim_{\epsilon\,\, \rightarrow\,\, 0}\;\, \Omega_{\; \pm\, ,\;\, \epsilon}^{\; (\, \lambda\, )\, }\, 
\nonumber\\
\quad\quad\quad\quad\quad\;\;  
=\,\, \exp\,\, (\,\, i\; e\;\, [\,\, a^{\, \dagger}\, (\, f_{\; v_{\, \mp}^{}\; \mathbf {x}}^{}\, )\, +
\, a\; (\, \overline {f\, }_{v_{\, \mp}^{}\; \mathbf {x}}^{}\, )\; ]\, )\; ,\;\,
\\
f_{\; v_{\, \mp}^{}\: \mathbf {x}}^{\; \mu}\; (\, \mathbf {k}\, ,\: t\, )\, =\,\, e^{\, -\, i\;\, \mathbf {k}\: \cdot\; \mathbf {x}}
\;\,\, \frac {\; \tilde {\rho\; }(\, \mathbf {k}\, )\;\, v_{\; \mp}^{\,\, \mu}\, } {\sqrt {\; 2\,\,\, \omega_{\,\, \mathbf {k}}^{}}
\; }\;\, \frac {\;\; i} {v_{\; \mp}^{}\cdot\, k\, }\, \cdot\quad\quad\quad
\end {eqnarray}
\end {theorem}
\begin {proof}
The proof is the same as that of Proposition \ref {proposizione Moller Feynman PFBR}.
\end {proof}
\ni
Scattering operators for the four-vector $\, BN\, $ model, interpreted as the result of soft photon corrections 
to an isometric scattering operator $\mathscr {W}\, $ in the velocity space, can be introduced by 
\begin {eqnarray}\label {matrice di scattering BN}
S_{\; \lambda\,\, }^{\,\, (FGB)\, }=\,\, \tau_{\; w}^{}-\lim_{\epsilon\,\, \rightarrow\; 0}\;
\lim_{t\, ,\,\, t\, '\; \rightarrow\, +\, \infty}\,\, \mathscr {U}_{\,\, I\; ,\,\, \lambda}^{\; (\, \epsilon\, )}
\; (\, t\, )\;\,\, \mathscr {W}\,\,\, \mathscr {U}_{\,\, I\; ,\,\, \lambda}^{\; (\, \epsilon\, )}\;
(\, t\; '\, )\, 
\nonumber\\
\equiv\,\, \tau_{\; w}^{}-\lim_{\epsilon\; \rightarrow\; 0}
\,\, S_{\; \lambda\,\, ,\,\, (\, \epsilon\, )\,\, }^{\,\, (FGB)}\, .
\end {eqnarray}
More generally, for the comparison of (\ref {matrice di scattering BN}) with the Feynman-Dyson expansion, 
the isometry $\, \mathscr {W}\, $ can be taken as acting on the entire space $ \mathscr {V}^{\; BN} $ and 
interpreted as yielding an infrared-finite $S\; $- matrix, obtained by factoring out from the scattering 
amplitudes the contributions given by the M\"{o}ller operators of the model.

In the remainder of this Section, we prove that the infrared diagrammatic of $\, QED\, $ is
reproduced with the help of the operators (\ref {Moller Feynman BN}). 
Regarding the infrared approximations and results within the perturbative-theoretic framework, we shall refer 
in the sequel to the streamlined treatment of \cite {Wei95,JR76}, while a more detailed analysis can be found 
in the classic work of Yennie, Frautschi and Suura \cite {YFS61}.

In what follows, we introduce an energy scale, say $\, \Lambda\, ,\, $ conventionally dividing the soft and
the hard photons; for photon canonical operator-valued distributions smeared by solutions of 
the wave equation (with mass $\, \lambda\, $) with energy below and above $\, \Lambda\, ,\, $ we shall
denote the corresponding subspaces of $\, \mathscr {F}\, $ respectively by $\, \mathscr {F}_{\, soft}^{}\, $ 
and $\, \mathscr {F}_{\; hard}^{}\, .$ 

For definiteness, we consider as a basic process $\, \alpha\rightarrow\beta\, $ the scattering of an electron by 
a potential and suppose that the incoming (outgoing) particle is described by a state of definite momentum, 
with four-velocity $\, v\; (\, v\: '\, )\, .\, $
The process is supposed not to involve low-energy photons, while it may involve hard photons.
Let $\, \eta_{\; e.\, m.\, }^{\,\, in}(\, \eta_{\; e.\, m.}^{\,\, out}\, )$ be the state of the incoming 
(outgoing) e.m. field; under the above assumptions, $\, \eta_{\; e.\, m.}^{\,\, in\, (\, out\, )}=
\Psi_{\, F\, }^{}\otimes\, \gamma_{\,\, e.\, m.}^{\,\, in\, (\, out\, )}\, ,$ with $\, \Psi_{\, F}^{}$ 
the vacuum vector of $\, \mathscr {F}_{\, soft}^{}\, $ and $\, \gamma_{\; e.\, m.}^{\,\, in}
(\, \gamma_{\; e.\, m.}^{\,\, out\, })\, $ belonging to $\, \mathscr {F}_{\; hard}^{}\, .\, $

The isometry $\, \mathscr {W}\, $ in (\ref {matrice di scattering BN}) is interpreted as the hard-photon 
part of the scattering operator, hence it is supposed to act as the identity operator on 
$\, \mathscr {F}_{\, soft\; }^{};\, $ moreover, according to the above interpretation of 
the models, the same property is assumed to hold for the restrictions of 
$\, \Omega_{\; \pm}^{\; (\, \lambda\, )}\, $ to $\, \mathscr {F}_{\; hard}^{}\, .\, $
The transition amplitude for the basic process $\, \alpha\rightarrow\beta\, $ is therefore
\begin {equation}\label {elemento di matrice base}
\mathscr {W}_{\,\, \beta\, \alpha}^{}\equiv\,\, \langle\,\, \psi_{\; v\: '}^{}\otimes\, \gamma_{\,\, e.\, m.\; }^{\,\, out},
\; \mathscr {W}\,\, (\; \psi_{\; v}^{}\otimes\, \gamma_{\,\, e.\, m.}^{\,\, in}\, )\, \rangle\; ,\quad\quad
\quad
\end {equation}
with $\, \psi_{\; v\, }^{}\, $ improper vectors describing a charged particle of four-velocity $\, v\, ,\, $ associated 
to the spectral resolution of the operators $\, \mathbf {v}\, .$\\
The matrix element for the same process, including the contributions due to the M\"{o}ller operators, is 
\begin {equation}\label {elemento di matrice totale}
(\; S_{\; \lambda\,\, }^{\,\, (FGB)}\; )_{\; \beta\; \alpha\, }^{}=\; \langle\; \Omega_{\; -}^{\: (\, \lambda\, )}
\; (\, \psi_{\; v\, '}^{}\otimes\, \eta_{\; e.\, m.}^{\,\, out}\, )\; ,\; \mathscr {W}\,\,\, 
\Omega_{\; +}^{\: (\, \lambda\, )}\; (\, \psi_{\; v}^{}\otimes\, 
\eta_{\; e.\, m.}^{\,\, in}\, )\, \rangle\; .
\end {equation}
The first result is given by:
\begin {theorem}
The soft-photon radiative corrections to the process $\alpha\rightarrow\beta$ are reproduced by 
the following contributions to the transition amplitude (\ref {elemento di matrice totale}):
\begin {equation}\label {correzioni radiative hamiltoniane}
\langle\; \Omega_{\; -\, ,\,\, v\, '}^{\: (\, \lambda\, )}\; \Psi_{\, F\; }^{},\, \Psi_{\, F\; }^{}\rangle
\,\, \langle\; \Psi_{\, F\,\, }^{},\; \Omega_{\; +\, ,\,\, v\;\, }^{\, (\, \lambda\, )}\Psi_{\, F\; }^{}\rangle
\;\, \exp\; (\, e^{\, 2}\,\, [\,\, a\, (\, \overline {f\, }_{v\, '\, \mathbf {x}}^{}\, )\, ,\, 
a^{\, \dagger}\, (\, f_{\: v\; \mathbf {x}}^{}\, )\, ]\, )\, .
\end {equation}
\end {theorem}
\begin {proof}
The proof is based on direct calculations.\\
The last term of (\ref {correzioni radiative hamiltoniane}) is related to ``virtual soft photons'' emitted 
from either the incoming or the outgoing external fermion leg and absorbed from the other, and is 
equal to 
\begin {equation}\label {correzioni radiative linee distinte}
\exp\,\, (\; e^{\, 2}\;\, v\, \cdot\, v\; '\, \int\,\, \frac {\, d^{\,\, 3\, }k} 
{2\;\, \omega_{\; \mathbf {k}\, }^{}}\;\,\, \tilde {\rho\,\, }^{2\,\, }
(\, \mathbf {k}\, )\;\,\,\frac {\;\;\, i} {\; v\, \cdot\, k}\;\, 
\frac {\;\;\;\; i} {\; v\; '\cdot\, k}\;\, )\; .
\quad
\end {equation}
The first (second) expression in brackets in (\ref {correzioni radiative hamiltoniane}) is related 
to the electron wave-function renormalization relative to the outgoing (incoming) electron
line; it is given by
\begin {eqnarray}\label {z non perturbativa}
\sqrt {\; Z_{\; 2\, ,\, IR\, ,\; \lambda\,\, }^{}(\, v\: '\, )\, }\, =
\,\, \exp\,\, (\; \frac {\, e^{\: 2}\,\, v^{\: 2}} {4\;\, }\,\, \int\,\, \frac {\, d^{\,\, 3\, }k} 
{\omega_{\,\, \mathbf {k}\, }^{\,\, 3}\, }\;\, \frac {\; \tilde {\rho\; }^{2\; }(\, \mathbf {k}\, )} 
{(\: 1-\hat {\mathbf {k}}\, \cdot\, \mathbf {v}\; )^{\; 2\, }}\,\,\, )\, 
\nonumber\\
=\,\, \exp\,\, (\, -\, \frac {\, e^{\: 2}} {2\, }\,\, \frac {\, \partial\,\,
\Sigma_{\, R\, ,\; \lambda\; }^{\, (\, \epsilon\, )}(\, v\: '\, )} 
{\partial\; (\, i\; \epsilon\, )\, }\,\, |_{\,\, \epsilon\; =\; 0}^{}\; )\: ,
\quad\quad\;\;
\end {eqnarray}
with $\, Z_{\; 2\, ,\: IR\, ,\; \lambda\,\, }^{}(\, v\, '\, )\, $ the infrared part of the wave-function renormalization 
constant relative to the outgoing leg and $\, \Sigma_{\, R\, ,\; \lambda\; }^{\: (\, \epsilon\, )}(\, v\: '\, )\, $ the 
self-energy insertion on the same line, the electron mass being renormalized:
\begin {eqnarray}\label {espansione epsilon autoenergia} 
\; i\,\, e^{\, 2}\,\,\, \Sigma_{\, R\, ,\; \lambda}^{\: (\, \epsilon\, )}\; (\, v\: '\, )\, =\;\, e^{\, 2}\,\, 
v^{\: 2}\;\, \epsilon\,\, \int\,\, \frac {d^{\,\, 3\, }k} {2\,\, \omega_{\; \mathbf {k}\, }^{\; 3}\, }
\;\, \frac {\tilde {\rho\; }^{2\; }(\, \mathbf {k}\, )} {(\, 1-\hat {\mathbf {k}}\, \cdot\, 
\mathbf {v}\, '\, )^{\,\, 2}\, }\, +\; O\; (\, \epsilon^{\; 2}\, )\; ,
\nonumber
\end {eqnarray}
with $\, \hat {\mathbf {k}}\equiv\, \mathbf {k}\, /\, \omega_{\; \mathbf {k}}^{}\, .\, $
Such a term is a residual infrared contribution, arising whenever ``on-shell'' mass-renormalization conditions 
are imposed \cite {JR76}. 

Summing (\ref {correzioni radiative linee distinte}) and (\ref {z non perturbativa}), the standard result
for the soft-photon radiative corrections (($13.2.4$),($13.2.5$) in \cite {Wei95}) is 
recovered.
\end {proof}
We can now examine the effects due to the emission of low-energy radiation.
For definiteness, we consider the contributions due to the emission of $\, n\, $ soft photons from the 
outgoing fermion line. 
Let $\, \Psi_{\; k_{\, 1}^{}\, ...\; k_{\, n\, }^{}}^{\,\, \mu_{\, 1}^{}\, ...\; \mu_{\, n}^{}}\, $ denote the vector
describing the $\, n\; $- photon state with polarization indices $\, \mu_{\; 1\, }^{}...\, \mu_{\: n}^{}$ 
and four-momenta $\, k_{\; 1\, }^{}...\, k_{\: n}^{}\, .\, $ 
It is straightforward to establish the following:
\begin {theorem}
The overall transition amplitude for the process $\alpha\rightarrow\beta$, with emission of $\, n\, $ soft 
photons described by the state $\, \Psi_{\: k_{\, 1}^{}\, ...\; k_{\, n\, }^{}}^{\,\, \mu_{\, 1}^{}\, ...\; \mu_{\, n}^{}}\, ,\, $
is reproduced by the matrix element
\begin {equation}
\langle\; \psi_{\; v\, '}^{}\otimes\, \Psi_{\, k_{\, 1}^{}\, ...\, k_{\, n\, }^{}}^{\; \mu_{\, 1}^{}\, ...\, \mu_{\, n}^{}\, }
\otimes\, \gamma_{\; e.\, m.\; }^{\,\, out},\; S_{\; \lambda\; ,\; (\, \epsilon\, )}^{\,\, (FGB)}\; (\, \psi_{\; v}^{}
\otimes\, \Psi_{\, F\, }^{}\otimes\, \gamma_{\; e.\, m.}^{\; in}\, )\, \rangle\, .
\end {equation}
\begin {proof}
A simple calculation yields
\begin {eqnarray}\label {correzione emissione fotoni} 
\langle\, \Psi_{\; k_{\, 1}^{}\, ...\; k_{\, n\, }^{}}^{\,\, \mu_{\, 1}^{}\, ...\; \mu_{\, n}^{}}\, ,\,\, 
\exp\; (\, -\, i\; e\;\, a^{\, \dagger}\, (\, f_{\: v\, '\: \mathbf {x}}^{\: (\, \epsilon\, )\,\, })\, )\;\, 
\Psi_{\, F\; }^{}\rangle
=\, \Pi_{\; j\; =\; 1}^{\; n}\; \frac {e\;\, \tilde {\rho\,\, }
(\, \mathbf {k}_{\, j}^{}\, )}{\sqrt {\;\, 2\;\, \omega_{\,\, \mathbf {k}_{\, j}^{}}^{}}\;\; }
\,\, \frac {\, {v\; '}^{\,\, \mu_{\, j}^{}}\,\,\, e^{\, -\, i\,\, \mathbf {k}\, \cdot\, \mathbf {x}_{\, j}^{}}}
{-\; v\; '\cdot\, k_{\; j}^{}-i\,\, \epsilon\, }\, \cdot
\nonumber\\
\end {eqnarray}
Recalling (\ref {Moller Feynman BN}), (\ref {matrice di scattering BN}) and (\ref {elemento di matrice totale}) 
and employing (\ref {correzione emissione fotoni}) one gets
\begin {eqnarray}
\; \langle\; \psi_{\; v\, '}^{}\otimes\, \Psi_{\, k_{\, 1}^{}\, ...\; k_{\, n\, }^{}}^{\; \mu_{\, 1}^{}\, ...\; \mu_{\, n}^{}\, }
\otimes\, \gamma_{\; e.\, m.\; }^{\,\, out},\; S_{\; \lambda\; ,\; (\, \epsilon\, )}^{\,\, (FGB)}\; (\, \psi_{\; v}^{}
\otimes\, \Psi_{\, F\, }^{}\otimes\, \gamma_{\; e.\, m.}^{\; in}\, )\, \rangle
\, 
\quad\quad\nonumber\\
=\, (\, S_{\; \lambda\; ,\; (\, \epsilon\, )}^{\,\, (FGB)}\, )_{\; \beta\; \alpha\, }^{}\,\, \Pi_{\; j\; =\; 1}^{\; n}\;\, \frac {\tilde {\rho\; }(\, \mathbf {k}_{\, j\, }^{})} 
{\sqrt {\;\, 2\;\, \omega_{\,\, \mathbf {k}_{\, j}^{}\, }^{}}\;\; }\, \sum_{l\;\, =\;\, 1\, }^{2}
\; \frac {\eta_{\,\, l}^{}\;\, e\;\, v_{\,\, l}^{\;\, \mu_{\, j}}}
{-\; v_{\,\, l}^{}\, \cdot\, k_{\; j}^{}-i\,\, \eta_{\,\, l}^{}\;\, 
\epsilon\, }\;\, ,
\nonumber
\\
v_ {\; 1\, }^{}\equiv\, v\, ,\, v_ {\; 2\, }^{}\equiv\, v\; '\, ,\; \eta_ {\; 2}^{}=\, 1\, =-\, \eta_ {\; 1}^{}\, ,
\nonumber
\end {eqnarray}
in agreement with the perturbative expression (($13.3.1$) in \cite {Wei95}).
\end {proof}
\end {theorem}
\ni
The infrared phases occurring in the transition amplitude for a process described by the sum of one-particle 
Hamiltonians, with at least two charged particles in either the initial or the final state, can also be recovered 
by explicit calculations. For brevity we do not report the details.

Finally we remark that the Bloch-Nordsieck expansion preserves the free-field character of 
$\, \partial\, \cdot\, A\, $ and consequently the gauge-invariance of the transition amplitudes, 
in the sense that terms from unphysical photon polarizations cancel.

\section* {Outlook}

\ni
By exploiting the control of the soft photon contributions to the Feynman-Dyson expansion 
of $\, QED\, ,$ gained through the four-vector Bloch-Nordsieck model, the recipes leading to 
infrared-finite inclusive cross-sections can also be formulated and discussed. 
In particular, it is possible to enlighten the non-perturbative limitations of such recipes, due to the 
superselection of particle momenta, taking advantage of the fact that within the hamiltonian
approach here developed issues connected with an order-by-order diagrammatic treatment 
are avoided.
We plan to report on these problems in a future paper.

\section* {Acknowledgments}

\ni
The ideas at the basis of this work mainly stem from a preliminary analysis of hamiltonian models 
by Giovanni Morchio and Franco Strocchi, devoted to a better understanding of the role of local and 
covariant formulations in the treatment of the infrared problem in Quantum Electrodynamics. 
I would like to thank Giovanni Morchio for extensive discussions on these topics and for a number of 
comments on the manuscript.


\begin {thebibliography} {50}

\bibitem [Ara81]{Ara81} A. Arai. Self-adjointness and spectrum of Hamiltonians in nonrelativistic quantum electrodynamics. \emph {J. Math. Phys.}, \textbf {22}(3):534--537, 1981.
\bibitem [Ara83]{Ara83} A. Arai. A note on scattering theory in non-relativistic quantum electrodynamics. \emph {J. Phys. \textbf {A}}: \emph {Math. Gen.}, \textbf {16}:49--70, 1983.
\bibitem  [Bla69]{Bla69} P. Blanchard. Discussion math\'ematique du mod\`ele de Pauli et Fierz relatif \`a la catastrofe infrarouge. \emph {Commun. Math. Phys.}, \textbf {19}(2):156--172, 1969.
\bibitem [Ble50]{Ble50} K. Bleuler. Eine neue Methode zur Behandlung der longitudinalen und skalaren photonen. \emph {Helv. Phys. Acta}, \textbf {23}:567--586, 1950.
\bibitem [BN37]{BN37} F. Bloch and A. Nordsieck. Note on the Radiation Field of the Electron. \emph {Phys. Rev.}, \textbf {52}:54--59, 1937.
\bibitem [Buc82]{Buc82} D. Buchholz. The Physical State Space of Quantum Electrodynamics. \emph {Commun. Math. Phys.}, \textbf {85}(1):49--71, 1982.
\bibitem [Buc86]{Buc86} D. Buchholz. Gauss' law and the infraparticle problem. \emph {Phys. Lett. \textbf {B}}, \textbf {174}(3):331--334, 1986.
\bibitem [Dys49]{Dys49} F. J. Dyson. The S Matrix in Quantum Electrodynamics. \emph {Phys. Rev.}, \textbf {75}:1736--1755, 1949.
\bibitem [Dys51]{Dys51} F. J. Dyson. Heisenberg Operators in Quantum Electrodynamics. I. \emph {Phys. Rev.}, \textbf {82}:428--439, 1951.
\bibitem [FL74]{FL74} W. G. Faris and R. B. Lavine. Commutators and Self-Adjointness of Hamiltonian Operators. \emph {Commun. Math. Phys.}, \textbf {35}:39--48, 1974.
\bibitem [FMS79a]{FMS79a} J. Fr\"{o}hlich, G. Morchio and F. Strocchi. Charged sectors and scattering states in quantum electrodynamics. \emph {Ann. Phys.}, \textbf {119}(2):241--284, 1979.
\bibitem [FMS79b]{FMS79b} J. Fr\"{o}hlich, G. Morchio and F. Strocchi. Infrared problem and spontaneous breaking of the Lorentz group in QED. \emph {Phys. Lett. \textbf {B}}, \textbf {89}(1):61--64, 1979.
\bibitem [FP38]{FP38} M. Fierz and W. Pauli. Zur Theorie der Emission langwelliger Lichtquanten. \emph {Nuovo Cimento}, \textbf {15}(3):167--188, 1938.
\bibitem [FPS74]{FPS74} R. Ferrari, L. E. Picasso and F. Strocchi. Some remarks on local operators in quantum electrodynamics. \emph {Commun. Math. Phys.}, \textbf {35}:25--38, 1974.
\bibitem [FPS77]{FPS77} R. Ferrari, L. E. Picasso and F. Strocchi. Local operators and charged states in quantum electrodynamics. \emph {Nuovo Cimento Soc. Ital. Fis.}, \textbf {A39}(1):1--8, 1977.
\bibitem [Gup50]{Gup50} S. N. Gupta. Theory of longitudinal photons in quantum electrodynamics. \emph {Proc. Phys. Soc. Lond. \textbf {A}}, \textbf {63}(7):681--691, 1950.
\bibitem [HS09]{HS09} F. Hiroshima and A. Suzuki. Physical State for Nonrelativistic Quantum Electrodynamics. \emph {Ann. Henri Poincar\'e}, \textbf {10}(5):913--953, 2009.
\bibitem [JR76]{JR76} J. M. Jauch and F. Rohrlich. \emph {The Theory of Photons and Electrons. Springer}, New York, second expanded edition, 1976.
\bibitem [Kat56]{Kat56} T. Kato. On linear differential equations in Banach spaces. \emph {Comm. Pure App. Math.}, \textbf {9}(3):479--486, 1956.
\bibitem [MS80]{MS80} G. Morchio and F. Strocchi. Infrared singularities, vacuum structure and pure phases in local quantum field theory. \emph {Ann. Henri Poincar\'e}, \textbf {33}(3):251--282, 1980.
\bibitem [MS83]{MS83} G. Morchio and F. Strocchi. A non-perturbative approach to the infrared problem in QED: Construction of charged states. \emph {Nucl. Phys. \textbf {B}}, 211:471--508, 1983.
\bibitem [Nai59]{Nai59} M. A. Naimark. \emph {Normed Rings}. Noordhoff, Groningen, 1959.
\bibitem [Nel64]{Nel64} E. Nelson. Interaction of Nonrelativistic Particles with a Quantized Scalar Field. \emph {J. Math. Phys.}, \textbf {5}(9):1190--1197, 1964.
\bibitem [RSII]{RS72a} M. Reed and B. Simon. \emph {Fourier Analysis, Self-Adjointness}, volume II of Methods of \emph {Modern Mathematical Physics}. Academic Press, 1972.
\bibitem [RSI]{RS72b} M. Reed and B. Simon. \emph {Functional Analysis}, volume I of Methods of \emph {Modern Mathematical Physics}. Academic Press, 1972.
\bibitem [Sch49]{Sch49} J. Schwinger. Quantum Electrodynamics. III. The Electromagnetic Properties of the Electron-Radiative Corrections to Scattering. \emph {Phys. Rev.}, \textbf {76}:790--817, 1949.
\bibitem [Steinb]{Ste00} O. Steinmann. \emph {Perturbative Quantum Electrodynamics and Axiomatic Field Theory}. Springer Verlag, New York, 2000.
\bibitem [St67]{St67} F. Strocchi. Gauge Problem in Quantum Field Theory. \emph {Phys. Rev.}, \textbf {162}:1429--1438, 1967.
\bibitem [SW74]{SW74} F. Strocchi and A. S. Wightman. Proof Of The Charge Superselection Rule In Local Relativistic Quantum Field Theory. \emph {J. Math. Phys.}, \textbf {15}:2198--2224, 1974.
\bibitem [Sym71]{Sym71} K. Symanzik. Lectures on lagrangian field theory. Desy report T-71/1, 1971.
\bibitem [Wei95]{Wei95} S. Weinberg. \emph {The Quantum Theory of Fields}, volume I. Cambridge University Press, Cambridge, 1995.
\bibitem [Wic50]{Wic50} G. C. Wick. The Evaluation of the Collision Matrix. \emph {Phys. Rev.}, \textbf {80}:268--272, 1950.
\bibitem [Wil67]{Wil67} R. M. Wilcox. Exponential Operators and Parameter Differentiation in Quantum Physics. \emph {J. Math. Phys.}, \textbf {8}:962--982, 1967.
\bibitem [YFS61]{YFS61} D. R. Yennie, S. C. Frautschi, and H. Suura. The Infrared Divergence Phenomena and High-Energy Processes. \emph {Ann. Phys.}, \textbf {13}:379--452, 1961.
\bibitem [PhDth]{Zer09} S. Zerella. \emph {Scattering Theories In Models Of Quantum Electrodynamics}. Ph.D. thesis, Universit\`a di Pisa, 2009. (unpublished).


\end {thebibliography}
\end {document}